\documentclass[journal,draftcls,onecolumn,12pt]{IEEEtran}

\usepackage{amsmath}
\usepackage{amssymb}
\usepackage[dvips]{graphicx}
\graphicspath{{./fig/}}
\usepackage{curves,color}
\usepackage{subfigure}
\usepackage{dsfont,slashbox,wasysym}
\usepackage{tikz}
\usepackage{pdfpages,enumerate}
\usepackage{multirow}
\usepackage{bm}
\usepackage{algorithmic}

\allowdisplaybreaks

\newtheorem{theorem}{Theorem}
\newtheorem{lemma}{Lemma}

\newtheorem{example}{Example}
\newtheorem{definition}{Definition}
\newtheorem{proposition}{Proposition}
\newtheorem{remark}{Remark}
\newtheorem{observation}{Observation}
\newtheorem{corollary}{Corollary}
\newtheorem{claim}{Claim}

\RequirePackage[normalem]{ulem} 
\RequirePackage{color}\definecolor{RED}{rgb}{1,0,0}\definecolor{BLUE}{rgb}{0,0,1} 

\newenvironment{mydescription}[1]
{\begin{list}{}%
{\renewcommand\makelabel[1]{##1\hfill}%
\settowidth\labelwidth{\makelabel{#1}}%
\setlength\leftmargin{\labelwidth}
\addtolength\leftmargin{\labelsep}}}
{\end{list}}

\begin{document}
\title{Equivalences Between Network Codes With Link Errors and Index Codes With Side Information Errors}
\author{Jae-Won~Kim~and~Jong-Seon~No, \IEEEmembership{Fellow,~IEEE}
\thanks{J.-W.~Kim and J.-S.~No are with the Department of Electrical and Computer Engineering, INMC, Seoul National University, Seoul 08826, Korea (e-mail: kjw702@ccl.snu.ac.kr, jsno@snu.ac.kr).}
}

\maketitle

\begin{abstract}

In this paper, new equivalence relationships between a network code with link errors (NCLE) and an index code with side information errors (ICSIE) are studied. First, for a given network coding instance, the equivalent index coding instance is derived, where an NCLE is converted to the corresponding ICSIE and vice versa. Next, for a given index coding instance, the equivalent network coding instance is also derived, where an ICSIE is converted to the corresponding NCLE and vice versa if a pair of encoding functions of an original link and the duplicated link are functionally related in the network code. Finally, several properties of an NCLE are derived from those of the equivalent ICSIE using the fact that the NCLE and the ICSIE are equivalent.  

\end{abstract}


\section{Introduction}\label{Introduction}

Network coding was introduced in \cite{networkflow} to improve the throughput gain of terminals in a network structure, where a source node transmits information to terminal nodes through links and internal nodes. In order to improve the throughput gain, some internal nodes encode their incoming symbols, which is called network coding. In \cite{linearnetworkcode}, it was proved that a linear network code for multicast in a network can achieve the max-flow bound. For multicast cases, there exist some algorithms to construct network codes achieving the maxflow-mincut capacity for a single source \cite{networkalgorithm1}, \cite{networkalgorithm2}. 

In contrast to an error-free link case, a network code dealing with erroneous data on links was also studied, referred to as a network code with link errors (NCLE) in this paper. As erroneous data on links in a network are considered, the number of overall link errors in a network which network codes can overcome was studied \cite{networkerror1}, \cite{networkerror2}. 

Index coding was introduced in \cite{Informed} for satellite communication systems which consist of one sender and several receivers. A sender has to transmit messages to receivers through a broadcast channel and receivers want to receive some messages and also know some messages priory as side information. Owing to its applications and relevance to other problems, index coding has attracted significant attention and various index coding schemes have accordingly been researched. For example, the optimal linear index coding scheme based on rank minimization over finite fields was introduced in \cite{ICSI} and random index coding was studied for infinitely long message length \cite{RI}.

In addition to researches on the index coding schemes, relevance to other problems has been researched such as the equivalence between network coding and index coding, topological interference management, and duality with distributed storage systems \cite{EQU}, \cite{TIM}, \cite{DSS}. There are also many researches on variations of index coding instances. For example, erroneous broadcast channels were considered in \cite{SECIC} and coded side information was studied in \cite{BCSI}, \cite{ICCSI}. Moreover, blind index coding instances where a sender only knows the probability distribution of side information were researched \cite{BlindIC} and functional index coding instances were introduced in \cite{FIC}. In contrast to conventional assumptions on side information, an index code in which side information errors exist, called an index code with side information errors (ICSIE) was studied in \cite{ICSIE}.

Among these researches, we focus on an equivalence between network coding and index coding \cite{EQU} in which their equivalence was introduced and a corresponding index coding instance was derived for a given network coding instance. It was also shown that any network codes can be converted to the corresponding index codes and vice versa. However, the equivalence between two problems for a given index coding instance was not presented in \cite{EQU}. In \cite{FEQU}, they showed an equivalence between network computation and functional index coding for a given network coding instance and also suggested their relation for a given index coding instance with the corresponding models for both a network coding instance and an index coding instance, called the equivalent index coding instance and network coding instance, respectively. However, their models of the corresponding instances are defined in a different manner, that is, if a given network coding instance is converted to the corresponding index coding instance and converted back to the network coding instance again, the re-converted network coding instance differs from the originally given network coding instance. Similarly, the same problem occurs to a given index coding instance. Thus, we propose a method to solve these problems in this paper.

In this paper, we show new equivalences between an NCLE and an ICSIE for both a given network coding instance and a given index coding instance. For a given network coding instance, the corresponding index coding instance is derived in a manner similar to that in an earlier study \cite{EQU} and convertibility of their solutions is proved. For a given index coding instance, we modify a given side information graph by adding receivers, messages, and edges or by deleting some edges in order to derive the corresponding network coding instance in a similar manner. We also show the convertibility of their solutions if a pair of encoding functions of an original link and the duplicated link are functionally related in the network code. Our models of the corresponding instances not only offer convertibility of the coding solutions but also ensure that a given network (index) coding instance is identical to the re-converted network (index) coding instance from the corresponding index (network) coding instance. Moreover, the equivalent index coding instance of a given network coding instance does not contain the receiver $\hat{t}_{\rm all}$, which was given in the earlier studies \cite{EQU}, \cite{FEQU}. In \cite{SEQU}, it was noted that an equivalence between secure network and index coding can be achieved without $\hat{t}_{\rm all}$. Similarly, we prove in detail that the receiver $\hat{t}_{\rm all}$ of the corresponding index coding instance is redundant in general. Since an NCLE and an ICSIE are equivalent, we derive several properties of an NCLE from the properties of an ICSIE such as the property of redundant links and the relationship between the conventional network code with error-free links and an NCLE.

The paper is organized as follows. Several definitions, notations, and problem settings are given in Section \ref{Preliminary}. The main results on equivalence relationships between an NCLE and an ICSIE for both a given network coding instance and a given index coding instance are derived in Section \ref{Equivalence}. In Section \ref{Properties}, several properties of an NCLE are derived from those of an ICSIE based on the equivalence between an NCLE and an ICSIE. Finally, conclusions are presented in Section \ref{Sec:Conclusion}.

\section{Preliminary}\label{Preliminary}

In this section, we define network codes with link errors and index codes with side information errors and then state their problem settings and notations, where hatted notations are used for index coding to avoid confusion.

\subsection{Notations}

Some of the notations are defined as follows:
\begin{enumerate}
\item $Z[n]$ denotes a set of positive integers $\{1,2,...,n\}$.
\item Let $\mathbb{F}_q$ be the finite field of size $q$, where $q$ is a power of prime and $\mathbb{F}_q^{*}=\mathbb{F}_q\setminus\{0\}$.
\item For the vector $\bold{X}\in \mathbb{F}_q^n$, $\rm{\rm{wt}}(\bold{X})$ denotes Hamming weight of $\bold{X}$.
\item Let $\bold{X}_D$ be a sub-vector $(X_{i_1},X_{i_2},\ldots,X_{i_{|D|}})$ of a vector $\bold{X}=(X_1,X_2,\ldots,X_n)\in \mathbb{F}_q^n$ for a subset $D=\{i_1,i_2,\ldots,i_{|D|}\}\subseteq Z[n]$, where $i_1<i_2<\ldots<i_{|D|}$.
\end{enumerate}

\subsection{Network Codes With Link Errors}
In this paper, in order to provide an equivalence between an NCLE and an ICSIE for any given index coding instance, a generalized network coding scenario is considered, where each internal node can resolve their erroneous incoming symbols.

For this scenario, if we know the probability distribution of the link errors, the throughput gain can be improved by assigning suitable error resistance capabilities to the internal nodes in a network structure. That is, large error resistance capabilities of internal nodes for the vulnerable links can improve throughput gain of an entire network because error propagation may be moderated. In this perspective, we introduce a new network code which deals with erroneous data on links.

First, we introduce a network coding instance with a network structure ${\mathbb{G}}=(V,E,\mathcal{F})$, where $V$ and $E$ denote the sets of nodes and edges in $\mathbb{G}$, respectively and a vector of the error resistance capabilities $\bm{\delta}$ described by a directed acyclic graph and a function of terminals $\mathcal{F}$ as follows:

\vspace{2mm}
\begin{enumerate}
\item $\bar{S}\subseteq V$ denotes a set of source nodes in ${\mathbb{G}}$, where source nodes do not have incoming links.
\item $S$ denotes a set of source messages, that is, $\bar{s}\in\bar{S}$ has some elements $s\in S$.
\item $T\subseteq V$ denotes a set of terminal nodes in ${\mathbb{G}}$, where terminal nodes do not have outgoing links.
\item $\mathcal{F}$ denotes a function of the terminal nodes in ${\mathbb{G}}$, which indicates a set of indices of each terminal's desired messages.
\item For a link $e=(u,v)\in E$, ${\rm In}(e)$ denotes a set of incoming links of $u$, where $u,v\in V$.
\item In the case of $u\in \bar{S}$, ${\rm In}(e)$ denotes a set of messages that $u$ has and ${\rm In}(t)$ denotes a set of incoming links of $t$ for which $t\in T$.
\item At the ends of the links, errors may occur due to transmissions through links, referred to as link errors and source nodes may have erroneous source symbols.
\item $\bm{\delta}=(\delta_{e_1},...,\delta_{e_{|E|}},\delta_{t_1},...,\delta_{t_{|T|}})$ is a vector whose elements correspond to the error resistance capability for each outgoing link from the node and terminal in $E\cup T$.
\item When it is straightforward, we regard $s$, $e$, and $t$ as some indices.
\end{enumerate}
\vspace{2mm}

In this network coding instance, we assume the followings:
\vspace{2mm}
\begin{enumerate}
\item Each message is one symbol in $\mathbb{F}_q$.
\item Each link carries one symbol in $\mathbb{F}_q$.
\item $X_s$ denotes an element of a message vector $\bold{X}\in \mathbb{F}_q^{|S|}$.
\item $X_e$ denotes a symbol on a link $e$ for which $e\in E$.
\item For a set $A\subseteq S$, $\bold{X}_A$ denotes a sub-vector of $\bold{X}$ and for a set $B\subseteq E$, $\bold{X}_B$ denotes a vector consisting of $|B|$ symbols of the corresponding links.
\item $\bold{\tilde{X}}_A=(\tilde{X}_1,...,\tilde{X}_{|A|})$ and $\bold{\tilde{X}}_B=(\tilde{X}_1,...,\tilde{X}_{|B|})$ denote vectors with erroneous symbol elements.
\end{enumerate}
\vspace{2mm}

\begin{figure}[t]
\centering
\includegraphics[scale=0.8]{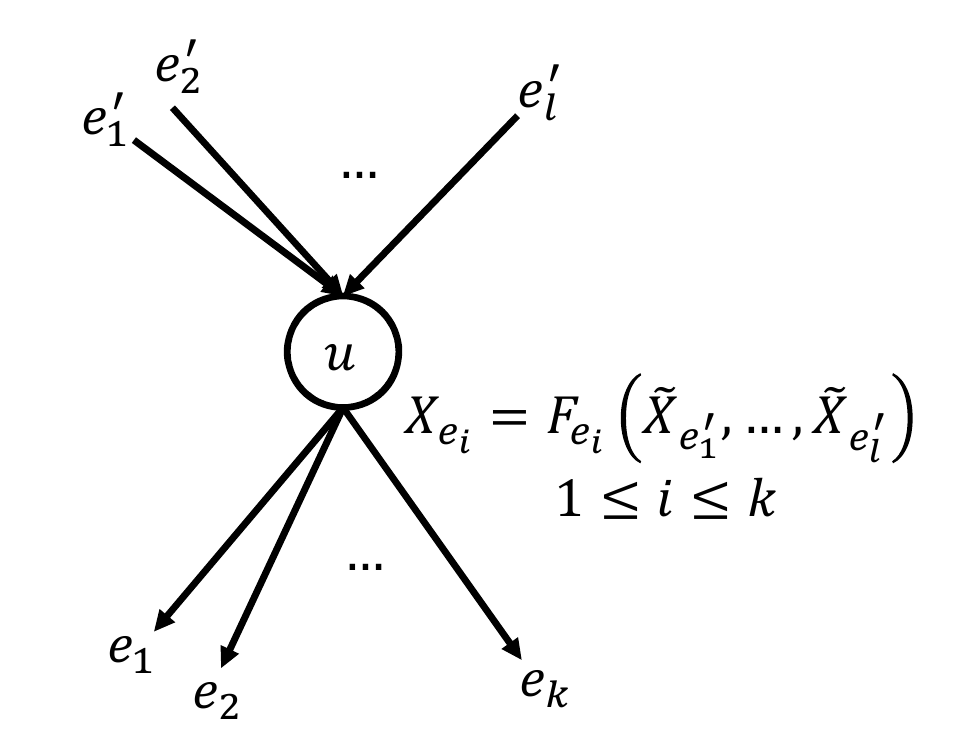}
\caption{Node processing in the network code.} 
\label{fig:node processing}
\end{figure}

Next, we describe node processing in the network code as in Fig. \ref{fig:node processing}, where $e^{\prime}_i, 1\leq i\leq l,$ denote the incoming edges of a node $u$ and $e_i, 1\leq i\leq k,$ denote the outgoing edges of $u$. At the node $u$, outgoing symbols for edges $e_i, 1\leq i\leq k,$ are computed by encoding functions as $X_{e_i}=F_{e_i}(\tilde{X}_{e^\prime_1},...,\tilde{X}_{e^\prime_l}), 1\leq i \leq k$. We consider a network code capable of resolving some link errors. Assume that there are less than or equal to $\delta_{e_1}$ symbol errors in the incoming links of $u$. If an encoding function $F_{e_1}$ can make a correct encoded outgoing symbol $X_{e_1}$ from $l$ incoming symbols with less than or equal to $\delta_{e_1}$ symbol errors, then $F_{e_1}$ is said to have an error resistance capability $\delta_{e_1}$. When $u$ is a source node, incoming symbols of $u$ denote the source messages possessed by $u$, meaning that up to $\delta_{e_1}$ message symbols are erroneous. Similarly, the decoding function $D_t$ of a terminal $t\in T$ is said to have an error resistance capability $\delta_t$ if $D_t$ can correctly obtain a decoded vector $\bold{X}_{\mathcal{F}(t)}$ from incoming symbols with less than or equal to $\delta_t$ symbol errors. In such a case, a network code with link errors is summarized as follows.

\vspace{2mm}
\begin{definition}
Let $\bm{\delta}=(\delta_{e_1},...,\delta_{e_{|E|}},\delta_{t_1},...,\delta_{t_{|T|}})$ be a vector whose elements correspond to the error resistance capability for each outgoing link and terminal in $E\cup T$. Then, a network code with link errors with parameters $(\bm{\delta},\mathbb{G})$ over $\mathbb{F}_q$, denoted by a $(\bm{\delta},\mathbb{G})$-NCLE consists of:

\begin{enumerate}
\item An encoding function $F_e: \mathbb{F}_q^{|{\rm In}(e)|}\rightarrow \mathbb{F}_q$ for $e\in E$
\item A decoding function $D_t: \mathbb{F}_q^{|{\rm In}(t)|}\rightarrow \mathbb{F}_q^{|\mathcal{F}(t)|}$ for $t\in T$

\item Satisfying $F_e(\bold{X}_{{\rm In}(e)})=F_e(\bold{\tilde{X}}_{{\rm In}(e)})$ and $D_t(\bold{X}_{{\rm In}(t)})=D_t(\bold{\tilde{X}}_{{\rm In}(t)})$ for any $e\in E$ and $t\in T$, where ${\rm wt}(\bold{X}_{{\rm In}(e)}-\bold{\tilde{X}}_{{\rm In}(e)})\leq\delta_e$ and ${\rm{wt}}(\bold{X}_{{\rm In}(t)}-\bold{\tilde{X}}_{{\rm In}(t)})\leq\delta_t$.
\end{enumerate}
\end{definition}
\vspace{2mm}

Note that the error resistance capabilities are defined for encoding and decoding functions, that is, encoding functions for the outgoing links of one node can have different error resistance capabilities despite the fact that they have identical erroneous incoming symbols. The above encoding functions of links are local functions. However, the global function $\bar{F}_e$ is defined as $\bar{F}_e(\bold{\tilde{X}}_S)=F_e(\bold{\tilde{X}}_{{\rm In}(e)})$.

\subsection{Index Codes With Side Information Errors}

We introduce index codes with side information errors as in \cite{ICSIE}. First, an index coding instance is described as follows:

\vspace{2mm}
\begin{enumerate}
\item There are one sender which has $n$ information messages as $\bold{\hat{X}}=(\hat{X}_1,\ldots,\hat{X}_n)\in\mathbb{F}_q^n$ and $m$ receivers (or users) $R_1,R_2,\ldots,R_m$, having sub-vectors of $\bold{\hat{X}}$ as side information.
\item Let ${{\mathcal{X}}_i}$ be the set of side information indices of a receiver $R_i$ for $i\in Z[m]$.
\item Each receiver $R_i$ wants to receive some elements in $\bold{\hat{X}}$, referred to as the wanted messages denoted by $\bold{\hat{X}}_{f(i)}$, where $f(i)$ represents the set of indices of the wanted messages of $R_i$ and $f(i)\cap{{\mathcal{X}}_i}=\phi$. 
\item A side information graph $\mathcal{G}$ shows the wanted messages and side information of all receivers and the sender knows $\mathcal{G}$. A side information graph is a bipartite graph which consists of message nodes and receiver nodes. A directed edge from a message node to a receiver node means that the receiver wants to receive that message. Conversely, a directed edge from a receiver node to a message node means that the receiver has that message as side information.
\item Let $\bm{\delta}_s=(\delta_s^{(1)},...,\delta_s^{(m)})$ be a vector whose elements correspond to the side information error resistance capability of each receiver.
\item The sender transmits messages to receivers through an error-free broadcast channel.

\end{enumerate}
\vspace{2mm}

Next, a $(\bm{\delta}_s,\mathcal{G})$-index code with side information errors is introduced. We consider an index code which can overcome arbitrary side information errors for each receiver, where each receiver does not know which side information is erroneous. Specifically, each receiver $R_i$ has a side information error resistance capability $\delta_s^{(i)}$ such that the receiver can decode the wanted messages even though less than or equal to $\delta_s^{(i)}$ symbols of side information are erroneous. Then, the $(\bm{\delta}_s,\mathcal{G})$-index code with side information errors is described as follows \cite{ICSIE}.

\vspace{2mm}
\begin{definition}
Let $\bm{\delta}_s=(\delta_s^{(1)},...,\delta_s^{(m)})$ be the vector of side information error resistance capabilities.
An index code with side information errors with parameters $(\bm{\delta}_s,\mathcal{G})$ over $\mathbb{F}_q$, denoted by a $(\bm{\delta}_s,\mathcal{G})$-ICSIE is a set of codewords having:
\vspace{2mm}
\begin{enumerate}
\item An encoding function $\hat{F}: \mathbb{F}_q^n\rightarrow\mathbb{F}_q^N$
\item A set of decoding functions $\hat{D}_1,\hat{D}_2,\ldots,\hat{D}_m$ such that $\hat{D}_i: {\mathbb{F}_q^N}\times{\mathbb{F}_q^{|{{\mathcal{X}}_i}|}}\rightarrow\mathbb{F}_q^{|f(i)|}$ satisfying
\begin{equation}
\hat{D}_i(\hat{F}(\bold{\hat{X}}),{\bold{\hat{X}}}_{\tilde{{\mathcal{X}}}_i})=\bold{\hat{X}}_{f(i)}
\nonumber
\end{equation}
for all $i\in Z[m]$, $\bold{\hat{X}}\in\mathbb{F}_q^n$, and ${\rm wt}(\bold{\hat{X}}_{{\mathcal{X}}_i}-\bold{\hat{X}}_{\tilde{{\mathcal{X}}}_i})\leq\delta_s^{(i)}$, where $\bold{\hat{X}}_{\tilde{{\mathcal{X}}}_i}=(\hat{X}_{\tilde{1}},...,\hat{X}_{\tilde{|\mathcal{X}_i|}})$ is the erroneous side information vector of a receiver $R_i$.  
\end{enumerate}
\end{definition}

\vspace{2mm}

A $(\bm{\delta}_s,\mathcal{G})$-ICSIE in \cite{ICSIE} is a linear code. However, a general index code containing a nonlinear case is considered in this paper. Thus, we should modify and re-prove some of the properties of a $(\bm{\delta}_s,\mathcal{G})$-ICSIE.

Let $\mathcal{I}(q,\mathcal{G},\bm{\delta}_s)$ be a set of vectors defined by
\begin{equation}
\mathcal{I}(q,\mathcal{G},\bm{\delta}_s)=\bigcup_{i\in Z[m]}\mathcal{I}_i(q,\mathcal{G},\delta_s^{(i)})
\nonumber
\end{equation}
where $\mathcal{I}_i(q,\mathcal{G},\delta_s^{(i)})=\{\bold{\hat{Z}}\in\mathbb{F}_q^n | {\rm wt}(\bold{\hat{Z}}_{\mathcal{X}_i})\leq2\delta_s^{(i)}, \bold{\hat{Z}}_{f(i)}\neq\bold{0}\}$.

Then, a property of a $(\bm{\delta}_s,\mathcal{G})$-ICSIE is given in the following theorem.
\vspace{2mm}
\begin{theorem}
A $(\bm{\delta}_s,\mathcal{G})$-ICSIE is valid if and only if 
\begin{equation}
\hat{F}(\bold{\hat{X}})\neq \hat{F}(\bold{\hat{X}}^{\prime}) \textrm{ for all } \bold{\hat{X}}-\bold{\hat{X}}^{\prime}\in\mathcal{I}(q,\mathcal{G},\bm{\delta}_s).
\label{equ:encoding}
\end{equation}
\label{encoding}

\begin{proof}
Each receiver $R_i$ has to recover $\bold{\hat{X}}_{f(i)}$ using the received codeword $\hat{F}(\bold{\hat{X}})$ and the side information $\bold{\hat{X}}_{\tilde{\mathcal{X}}_i}$. Then, the sender has to encode some confusing messages as different codewords. Because each receiver $R_i$ is only interested in $\bold{\hat{X}}_{f(i)}$, the codewords of distinct messages with an identical $\bold{\hat{X}}_{f(i)}$ do not need to be distinguished. Moreover, the codewords of two messages $\bold{\hat{X}}$ and $\bold{\hat{X}}^{\prime}$ such that ${\rm{wt}}(\bold{\hat{X}}_{\mathcal{X}_i}-\bold{\hat{X}}^{\prime}_{\mathcal{X}_i})>2\delta_s^{(i)}$ do not need to be distinguished because they can be distinguished by the side information of $R_i$. Thus, only problematic types of messages can be represented by two messages $\bold{\hat{X}}$ and $\bold{\hat{X}}^{\prime}$ such that $\bold{\hat{X}}_{f(i)}\neq\bold{\hat{X}}^{\prime}_{f(i)}$ and ${\rm{wt}}(\bold{\hat{X}}_{\mathcal{X}_i}-\bold{\hat{X}}^{\prime}_{\mathcal{X}_i})\leq2\delta_s^{(i)}$. Given that $\bold{\hat{X}}$ and $\bold{\hat{X}}^{\prime}$ are confusing, $\hat{F}(\bold{\hat{X}})\neq\hat{F}(\bold{\hat{X}}^{\prime})$ should be satisfied for all $i\in Z[m]$. 
\end{proof}
\end{theorem}
\vspace{2mm}

\begin{remark}

For a linear $(\bm{\delta}_s,\mathcal{G})$-ICSIE, (\ref{equ:encoding}) becomes $\bold{\hat{Z}}G\neq\bold{0}$ for all $\bold{\hat{Z}}\in\mathcal{I}(q,\mathcal{G},\bm{\delta}_s)$, where $G$ is the corresponding generator matrix.
\end{remark}
\vspace{2mm}

Let $\Phi$ be a set of subsets of $Z[n]$ defined by
\begin{equation}
\Phi=\{B\subseteq Z[n]\big||\mathcal{X}_i\cap B|\geq2\delta_s^{(i)}+1 \textrm{ for all } i\in Z[m] \textrm{ s.t. } f(i)\cap B\neq \phi\}
\nonumber
\end{equation}
for the side information graph $\mathcal{G}$ of a $(\bm{\delta}_s,\mathcal{G})$-ICSIE. Then, we have the following definition for a $\bm{\delta}_s$-cycle.

\vspace{2mm}
\begin{definition}
For a $(\bm{\delta}_s,\mathcal{G})$-ICSIE, a subgraph $\mathcal{G}^{\prime}$ of $\mathcal{G}$ is termed a $\bm{\delta}_s$-cycle if the set of message node indices of $\mathcal{G}^{\prime}$ is an element of $\Phi$ (i.e., $B$) and the set of user node indices of $\mathcal{G}^{\prime}$ consists of $i\in Z[m]$ such that $f(i)\in B$ and its edges consist of the corresponding edges in $\mathcal{G}$. The graph $\mathcal{G}$ is said to be $\bm{\delta}_s$-acyclic if there is no $\bm{\delta}_s$-cycle in $\mathcal{G}$.
\label{def:delta cycle}
\end{definition} 
\vspace{2mm}

A $\bm{\delta}_s$-cycle is an important subgraph for a $(\bm{\delta}_s,\mathcal{G})$-ICSIE problem because the existence of the $\bm{\delta}_s$-cycle is a necessary and sufficient condition for the possibility to reduce its codelength. The following lemma shows the importance of the $\bm{\delta}_s$-cycle.

\vspace{2mm}
\begin{lemma}
$\mathcal{G}$ is $\bm{\delta}_s$-acyclic if and only if $N_{\rm opt}^q(\bm{\delta}_s,\mathcal{G})=n$ for a $(\bm{\delta}_s,\mathcal{G})$-ICSIE, where $N_{\rm opt}^q(\bm{\delta}_s,\mathcal{G})$ is the optimal codelength.
\begin{proof}
The sufficiency part is similarly proved as in \cite{ICSIE} by showing the linear index code with codelength $n-1$, that is, $(\hat{X}_1+\hat{X}_2,\hat{X}_2+\hat{X}_3,...,\hat{X}_{n-1}+\hat{X}_n)$. The necessity part is based on the fact that $\mathcal{I}(q,\mathcal{G},\bm{\delta}_s)$ is a set of all vectors in $\mathbb{F}_q^n$ except for $\bold{0}$ if $\mathcal{G}$ is $\bm{\delta}_s$-acyclic. Specifically, because $Z[n]$ is not a $\bm{\delta}_s$-cycle, we can assume that there is at least one receiver $R_1$ with a wanted message $\hat{X}_1$ and $|\mathcal{X}_1\cap Z[n]|\leq 2\delta_s^{(1)}$ without loss of generality. Then, every $\bold{\hat{Z}}\in\mathbb{F}_q^n$ with $\hat{Z}_1\neq 0$ is included in $\mathcal{I}(q,\mathcal{G},\bm{\delta}_s)$. Similarly, because $Z[n]\setminus\{1\}$ is not a $\bm{\delta}_s$-cycle, we can assume that there is at least one receiver $R_2$ with a wanted message $\hat{X}_2$ and $|\mathcal{X}_2\cap \{Z[n]\setminus\{1\}\}|\leq 2\delta_s^{(2)}$. Then, every $\bold{\hat{Z}}\in\mathbb{F}_q^n$ with $\hat{Z}_1=0$ and $\hat{Z}_2\neq 0$ is included in $\mathcal{I}(q,\mathcal{G},\bm{\delta}_s)$. The similar result for $R_i$ is that every $\bold{\hat{Z}}\in\mathbb{F}_q^n$ with $\hat{Z}_1=\hat{Z}_2=\cdots=\hat{Z}_{i-1}=0$ and $\hat{Z}_{i}\neq 0$ is included in $\mathcal{I}(q,\mathcal{G},\bm{\delta}_s)$, which means that $\mathcal{I}(q,\mathcal{G},\bm{\delta}_s)=\mathbb{F}_q^n\setminus\{\bold{0}\}$. In this case, all of the message vectors in $\mathbb{F}_q^n$ should be encoded to different codewords. Thus, $N_{\rm opt}^q(\bm{\delta}_s,\mathcal{G})=n$.
\end{proof}
\label{delta_s-cycle}
\end{lemma}
\vspace{2mm}
 
\section{Equivalences Between Network Codes With Link Errors and Index Codes With Side Information Errors }\label{Equivalence}

In this section, we prove the equivalences between network codes with link errors and index codes with side information errors. First, their equivalence is proved for a given network coding instance, similar to an earlier approach in \cite{EQU}. We also show some differences from that in \cite{EQU} for the corresponding index coding instance. Second, their equivalence is proved for a given index coding instance. For a given index coding instance, Gupta and Rajan defined the corresponding network coding instance and showed an equivalence between a network computation problem and a functional index coding problem \cite{FEQU}. However, the equivalence in \cite{FEQU} for a given index coding instance has some weak points, which will be explained in this section. In order to mitigate these weak points, for a given index coding instance, we introduce a corresponding network coding instance which differs from that in \cite{FEQU} and show a different equivalence relationship between them. In the following definition, the equivalence between two problems is described.

\vspace{2mm}
\begin{definition}
NCLE and ICSIE problems are said to be equivalent if and only if the NCLE can be converted to the corresponding ICSIE and vice versa. 
\end{definition}
\vspace{2mm}

\subsection{Equivalence for a Given Network Coding Instance}

For a given network coding instance, we can construct the corresponding index coding instance in a manner similar to that in the aforementioned research \cite{EQU}. The differences between our corresponding model and that in \cite{EQU} are the error resistance capabilities and the existence of the receiver $\hat{t}_{\rm all}$. In what follows, the relationship between two coding instances of a $(\bm{\delta},\mathbb{G})$-NCLE and the corresponding $(\bm{\delta}_s,\mathcal{G})$-ICSIE is given as follows:

\begin{enumerate}
\item A sender of a $(\bm{\delta}_s,\mathcal{G})$-ICSIE has a message $\bold{\hat{X}}=(\bold{\hat{X}}_S, \bold{\hat{X}}_E)$ and there are $|E|+|T|$ receivers, each of which is a corresponding receiver $R_e$ of a link or $R_t$ of a terminal in a given network coding instance.
\item For $e\in E$, $R_e$ of the $(\bm{\delta}_s,\mathcal{G})$-ICSIE can be described as $\mathcal{X}_e={\rm In}(e)$, $f(e)=\{e\}$, and $\delta_s^{(e)}=\delta_e$.
\item For $t\in T$, $R_t$ of the $(\bm{\delta}_s,\mathcal{G})$-ICSIE can be described as $\mathcal{X}_t={\rm In}(t)$, $f(t)=\mathcal{F}(t)$, and $\delta_s^{(t)}=\delta_t$.
\item The codelength of the $(\bm{\delta}_s,\mathcal{G})$-ICSIE is the number of links in a $(\bm{\delta},\mathbb{G})$-NCLE.
\end{enumerate}

\begin{figure}[t]
\centering
\subfigure[]{\includegraphics[scale=0.7]{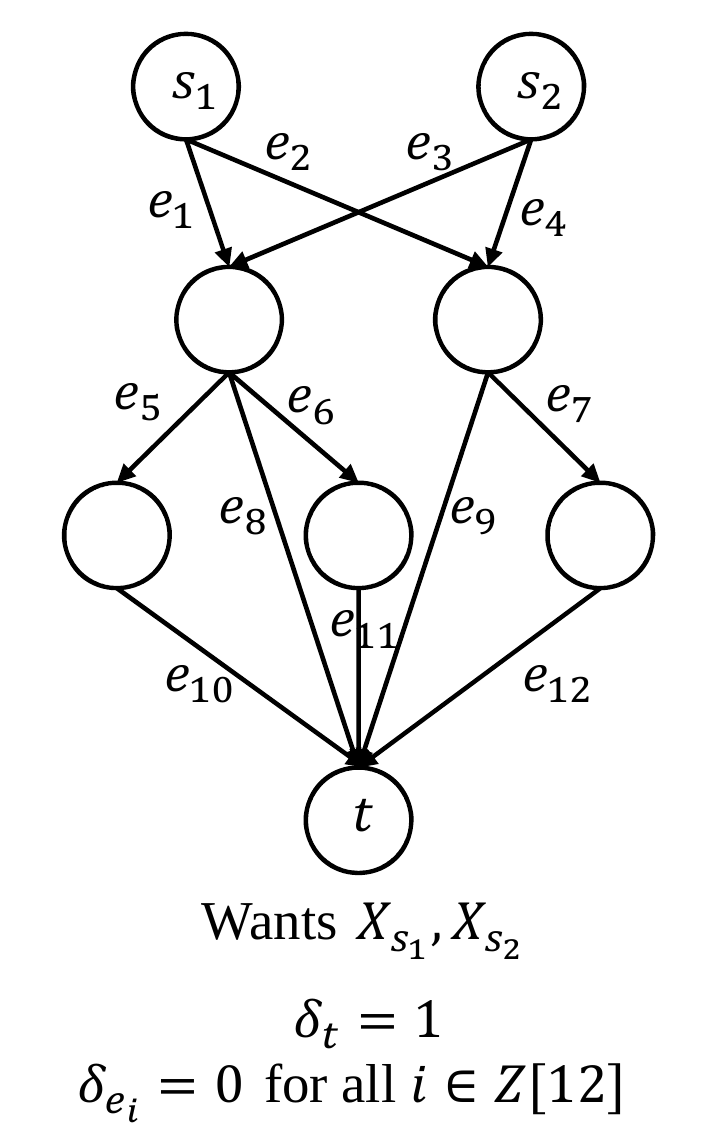}
\label{fig:fig1}}
\subfigure[]{\includegraphics[scale=0.7]{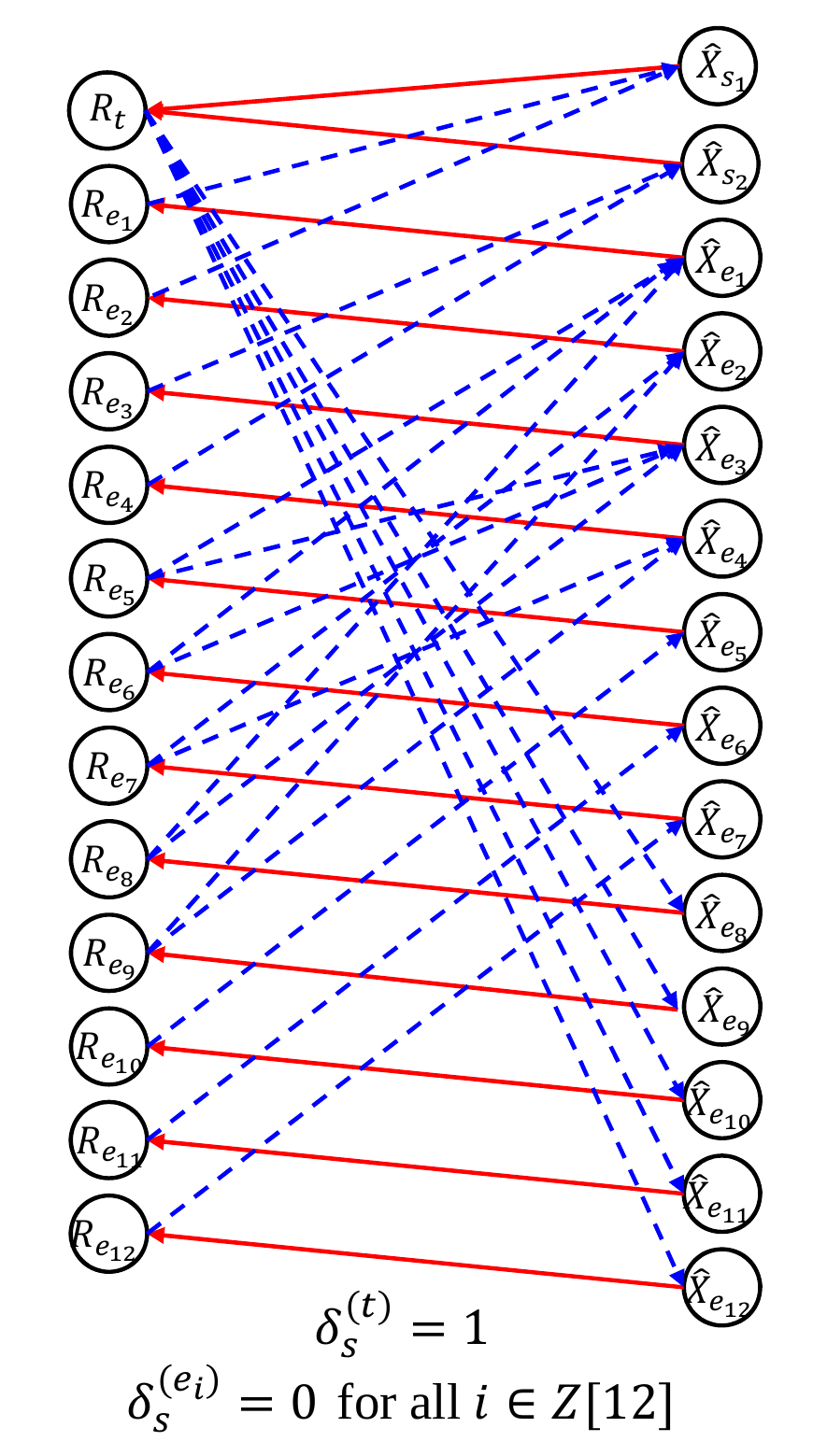}
\label{fig:fig7}}

\caption{An example of a network coding instance and the corresponding index coding instance. (a) A network coding instance (b) The corresponding side information graph with $\bm{\delta}_s$.} 
\label{fig:network example}
\end{figure}

This relationship is derived from a given network coding instance. Fig. \ref{fig:network example} shows an example of a network coding instance and the corresponding index coding instance. Before showing validity of this model, some restrictions for the corresponding index coding instance should be satisfied as shown in the following proposition.

\vspace{2mm}
\begin{proposition}
The corresponding index coding instance of a given network coding instance should satisfy the followings:
\begin{enumerate}
\item In the corresponding side information graph $\mathcal{G}$, there is no cycle in a subgraph which consists of $\{R_e | e\in E\}$ and $\{\hat{X}_e | e\in E\}$.
\item Each element of $\{\hat{X}_e | e\in E\}$ should be wanted by one receiver.
\item If $R_i$ has $\hat{X}_s$ as side information for $s\in S$, $R_i$ cannot have $\hat{X}_e$ as side information for $e\in E$.
\end{enumerate}

\begin{proof}
If a network structure is valid, the network structure is directed acyclic and the source nodes are not intermediate. Thus, to be directed acyclic, 1) should be satisfied. 2) is due to our setting of the relation and 3) should be satisfied because the source nodes are not intermediate nodes. 
\end{proof}
\label{prop:condition}
\end{proposition}
\vspace{2mm}

From Proposition \ref{prop:condition}, we note that the corresponding models do not cover all index coding instances but cover some index coding instances satisfying conditions in the above proposition necessarily. However, it is important to note that all network structures can be covered by this model. 

In this model, one difference from that in \cite{EQU} is the existence of the receiver $\hat{t}_{\rm all}$ which can be described as $\mathcal{X}_{\hat{t}_{\rm all}}=S$, $f(\hat{t}_{\rm all})=E$, and $\delta_s^{(\hat{t}_{\rm all})}=0$. In fact, the existence of $\hat{t}_{\rm all}$ in \cite{EQU} originates from a directed acyclic network structure. Thus, an identical result can be obtained even if we remove $\hat{t}_{\rm all}$ in the corresponding index coding instance as in the following proposition.

\vspace{2mm}
\begin{proposition}
For a given network coding instance, the modeling of the corresponding index coding instance in \cite{EQU} obtains an identical result even if the receiver $\hat{t}_{\rm all}$ is removed, that is, $\hat{t}_{\rm all}$ is redundant.

\begin{proof}
From 1) of Proposition \ref{prop:condition}, we can see that there is no cycle in a subgraph which consists of $\{R_e | e\in E\}$ and $\{\hat{X}_e | e\in E\}$ and thus there is no $\bm{\delta}_s$-cycle. Since this subgraph is $\bm{\delta}_s$-acyclic, the optimal codelength for the subgraph is $|E|$ by Lemma \ref{delta_s-cycle}, meaning that every vector $\bold{\hat{Z}}\in\mathbb{F}_q^{|S|+|E|}$ such that $\bold{\hat{Z}}_S=\bold{0}$ and $\bold{\hat{Z}}_E\neq\bold{0}$ belongs to $\mathcal{I}(q,\mathcal{G},\bm{\delta}_s)$. In \cite{EQU}, $\hat{t}_{\rm all}$ wants to receive $\bold{\hat{X}}_{E}$ and has $\bold{\hat{X}}_S$ as side information with $\delta_s^{(\hat{t}_{\rm all})}=0$. In fact, we do not need $\hat{t}_{\rm all}$ because there is no cycle in the subgraph mentioned above. From Theorem \ref{encoding}, $\mathcal{I}_{\hat{t}_{\rm all}}(q,\mathcal{G},\delta_s^{(\hat{t}_{\rm all})})$ is the set of all vectors in $\mathbb{F}_q^{|E|+|S|}$ such that $\bold{\hat{Z}}_S=\bold{0}$ and $\bold{\hat{Z}}_E\neq\bold{0}$. Since every vector in $\mathcal{I}_{\hat{t}_{\rm all}}(q,\mathcal{G},\delta_s^{(\hat{t}_{\rm all})})$ is already included in $\mathcal{I}(q,\mathcal{G},\bm{\delta}_s)$, the receiver $\hat{t}_{\rm all}$ can be removed from the corresponding index coding instance.
\end{proof}
\label{prop:tall}
\end{proposition}
\vspace{2mm}

Thus, we can remove the receiver $\hat{t}_{\rm all}$ from the corresponding index coding instance. From the proof of Proposition \ref{prop:tall} and Lemma \ref{delta_s-cycle}, the following observation is given.
\vspace{2mm}
\begin{observation}
The optimal index codelength of the corresponding index coding instance is larger than or equal to $|E|$. 
\label{ob:codelength}
\end{observation}
\vspace{2mm}

At this point, we prove validity of this model and the equivalence between an NCLE and an ICSIE. In order to show that they are equivalent, the following lemma is needed.

\vspace{2mm}
\begin{lemma}
In the equivalent $(\bm{\delta}_s,\mathcal{G})$-ICSIE for a given network coding instance, there is a unique $\bold{\hat{X}}_E$ such that $\hat{F}(\bold{\hat{X}}_S,\bold{\hat{X}}_E)=\bm{\sigma}$ for any codeword $\bm{\sigma}\in\mathbb{F}_q^{|E|}$ and $\bold{\hat{X}}_S$.
\begin{proof}
From Proposition \ref{prop:condition}, there is no cycle in a subgraph which consists of the set of receivers $\{R_e | e\in E\}$ and the set of messages $\{\hat{X}_e | e\in E\}$. Thus, $N_{\rm{opt}}^q(\bm{\delta}_s,\mathcal{G})\geq |E|$ and we assume that a $(\bm{\delta}_s,\mathcal{G})$-ICSIE with codelength $|E|$ exists. Since different symbols of $\bold{\hat{X}}_E$ for given $\bold{\hat{X}}_S$ result in different codewords and the codelength is $|E|$ by Lemma \ref{delta_s-cycle}, there exists unique $\bold{\hat{X}}_E$ such that $\hat{F}(\bold{\hat{X}}_S,\bold{\hat{X}}_E)=\bm{\sigma}$ for the above conditions.

\end{proof}
\label{lemma:existence}
\end{lemma}
\vspace{2mm}

Next, the equivalence between an NCLE and an ICSIE for a given network coding instance is shown in the following theorem. 

\vspace{2mm}
\begin{theorem}
For a given network coding instance, a $(\bm{\delta},\mathbb{G})$-NCLE exists if and only if the corresponding $(\bm{\delta}_s,\mathcal{G})$-ICSIE exists.
\begin{proof}
This can be proved by a method similar to that in \cite{EQU} but the differences are the existence of $\hat{t}_{\rm all}$ and the fact that there are link errors and side information errors.

Necessity: Assume that there exists a $(\bm{\delta},\mathbb{G})$-NCLE. First, the encoding function of the corresponding $(\bm{\delta}_s,\mathcal{G})$-ICSIE is defined as $\hat{F}(\bold{\hat{X}})=\bold{\hat{X}}_B=(\hat{X}_B(e): e\in E)$ such that $\hat{X}_B(e)=\hat{X}_e+\bar{F}_e(\hat{X}_1,...,\hat{X}_{|S|})$. Next, we define the decoding functions and show that all receivers in the corresponding index coding instance can recover what they want. 
It is already given that $R_e$ can be described as $\mathcal{X}_e={\rm In}(e)$, $f(e)=\{e\}$, and $\delta_s^{(e)}=\delta_e$. Thus, for each $e^{\prime}\in {\rm In}(e)$, the decoder can compute $\hat{X}_B(e^{\prime})-\hat{X}_{\tilde{e^{\prime}}}$, which can be an erroneous value of $\bar{F}_{e^{\prime}}(\hat{X}_1,...,\hat{X}_{|S|})$. Then, $\delta_s^{(e)}$ symbols of them can be erroneous. Since the link $e$ in the network coding instance has an error resistance capability $\delta_e=\delta_s^{(e)}$, evaluating these symbols with $F_e$ results in the correct value of $\bar{F}_e(\hat{X}_1,...,\hat{X}_{|S|})$. Now, $R_e$ can obtain $\hat{X}_e$ by subtracting $\bar{F}_e(\hat{X}_1,...,\hat{X}_{|S|})$ from $\hat{X}_B(e)$. It is also given that $R_t$ can be described as $\mathcal{X}_t={\rm In}(t)$, $f(t)=\mathcal{F}(t)$, and $\delta_s^{(t)}=\delta_t$. Similar to the $R_e$ case, $R_t$ can recover what it wants because it can obtain $\bar{F}_e(\hat{X}_1,...,\hat{X}_{|S|})$ for all $e\in {\rm In}(t)$, whose $\delta_s^{(t)}$ symbols can be erroneous. However, evaluating these symbols using the decoding function $D_t$ of the network coding instance results in the correct values because $\delta_t=\delta_s^{(t)}$.

Sufficiency: Assume that there exists a $(\bm{\delta}_s,\mathcal{G})$-ICSIE together with its codeword $\bm{\sigma}$ by Lemma \ref{lemma:existence}. Then, the encoding functions of the links and the decoding functions of the terminals in the corresponding network coding instance are defined using the decoding functions of the $(\bm{\delta}_s,\mathcal{G})$-ICSIE. For $e\in E$, $F_e$ is defined as a function whose output is $X_e=\hat{D}_e(\bm{\sigma},(\tilde{X}_{e^{\prime}} : e^{\prime}\in {\rm In}(e)))$. For $t\in T$, $D_t$ is defined as a function whose output is $\hat{D}_{t}(\bm{\sigma},(\tilde{X}_{e^{\prime}} : e^{\prime}\in {\rm In}(t)))$. Without loss of generality, we assume that $\bold{X}_S=\bold{\hat{X}}_S$ and then, there is a unique $\bold{\hat{X}}_E$ such that $\hat{F}(\bold{\hat{X}}_S,\bold{\hat{X}}_E)=\bm{\sigma}$ for any $\bold{\hat{X}}_S$ from Lemma \ref{lemma:existence}. For $e \in E$, $\hat{D}_e(\bm{\sigma},(\tilde{X}_{e^{\prime}} : e^{\prime}\in {\rm In}(e)))=\hat{X}_e$ because $\delta_e=\delta_s^{(e)}$ and $\hat{X}_e$ is unique. For $t\in T$, $\hat{D}_{t}(\bm{\sigma},(\tilde{X}_{e^{\prime}} : e^{\prime}\in {\rm In}(t)))=\bold{\hat{X}}_{\mathcal{F}(t)}=\bold{X}_{\mathcal{F}(t)}$ because $\delta_t=\delta_s^{(t)}$.
\end{proof}
\label{theorem:equ}
\end{theorem}
\vspace{2mm}

Thus, Theorem \ref{theorem:equ} tells us that an NCLE problem is equivalent to the corresponding ICSIE problem when a network coding instance is given. The following example shows the equivalence between the two problems.

\vspace{2mm}
\begin{example}

Suppose that a given network coding instance and the corresponding side information graph with $\bm{\delta}_s$ are given as in Fig. \ref{fig:network example}. A network code for Fig. \ref{fig:fig1} can be described as follows:
\begin{enumerate}
\item $X_{e_1}=X_{e_2}=X_{e_5}=X_{e_8}=X_{e_{10}}=X_{s_1}$
\item $X_{e_3}=X_{e_4}=X_{e_7}=X_{e_{12}}=X_{s_2}$
\item $X_{e_6}=X_{e_9}=X_{e_{11}}=X_{s_1}+X_{s_2}$
\item $D_t$ is given in Algorithm 1.
\end{enumerate}

\begin{figure}[t]
\noindent\fbox{\parbox{0.98\linewidth}{
\textbf{Algorithm 1: Decoding procedure for the terminal $t$} \\
\textbf{Input}: $\tilde{X}_{s_1}$, $\tilde{X}_{s_1}$, $\tilde{X}_{s_2}$, $\tilde{X}_{s_1}+\tilde{X}_{s_2}$, and  $\tilde{X}_{s_1}+\tilde{X}_{s_2}$, one of which may be erroneous.\\
\textbf{Output}: $X_{s_1}$ and $X_{s_2}$
\begin{mydescription}{Step 1)}
\item[\textbf{Step 1)}]  
Compare two received symbols of $\tilde{X}_{s_1}+\tilde{X}_{s_2}$.
\item[\textbf{Step 2)}] 
If they are different, determine the received $\tilde{X}_{s_1}$ and $\tilde{X}_{s_2}$ as outputs and skip the following steps.
\item[\textbf{Step 3)}] 
If they are identical, compare two received symbols of $\tilde{X}_{s_1}$.
\item[\textbf{Step 4)}] 
If two received symbols of $\tilde{X}_{s_1}$ are identical, determine the received $\tilde{X}_{s_1}$ as $X_{s_1}$ and $X_{s_2}$ can be obtained from $\tilde{X}_{s_1}+\tilde{X}_{s_2}-\tilde{X}_{s_1}$.
\item[\textbf{Step 5)}]
If two received symbols of $\tilde{X}_{s_1}$ are different, determine the received $\tilde{X}_{s_2}$ as $X_{s_2}$ and $X_{s_1}$ can be obtained from $\tilde{X}_{s_1}+\tilde{X}_{s_2}-\tilde{X}_{s_2}$. 
\end{mydescription}
}}
\end{figure}

Then, the corresponding index code for Fig. \ref{fig:fig7} can be described as follows:
\begin{enumerate}
\item The transmitted codeword $\hat{F}(\bold{\hat{X}})=\hat{{\bold{X}}}_B=(\hat{X}_B(e) : e\in E)$ consists of 12 components.
\item $\hat{X}_B(e_i)=\hat{X}_{e_i}+\hat{X}_{s_1}$ for $i\in \{1, 2, 5, 8, 10\}$
\item $\hat{X}_B(e_i)=\hat{X}_{e_i}+\hat{X}_{s_2}$ for $i\in \{3, 4, 7, 12\}$
\item $\hat{X}_B(e_i)=\hat{X}_{e_i}+\hat{X}_{s_1}+\hat{X}_{s_2}$ for $i\in \{6, 9, 11\}$
\item The decoding functions of receivers can be defined as in Theorem \ref{theorem:equ}.
\end{enumerate}

Thus, by finding a network code for a given network coding instance, we can find the corresponding index code. Furthermore, if we find an index code with codelength $|E|$ for Fig. \ref{fig:fig7}, we can obtain the corresponding network code for Fig. \ref{fig:fig1}.

\end{example}
\vspace{2mm}

\subsection{Equivalence for a Given Index Coding Instance}

Similarly, for a given index coding instance, we can construct the corresponding network coding instance. Since some index coding instances cannot be converted to corresponding network coding instances, as noted in Proposition $\ref{prop:condition}$, it is necessary to modify a given index coding instance to use the previous relationship between two coding instances. Specifically, some receivers and messages are added to the given index coding instance, where $\mathcal{G}$ becomes $\mathcal{G}^{\prime}$ and $\bm{\delta}_s$ becomes $\bm{\delta}_s^{\prime}$.

For simplicity, we can assume that every receiver wants to receive only one message because a receiver who wants to receive more than one message can be split into receivers with identical side information. Now, we explain how to make $\mathcal{G}^{\prime}$, $\bm{\delta}_s^{\prime}$, and the corresponding network coding instance including $\mathbb{G}$ and $\bm{\delta}$.

In order to make the corresponding network coding instance using the same relationship between two coding instances in the previous section, it is necessary to determine which part is $\bold{\hat{X}}_E$ or $\bold{\hat{X}}_S$ for a given side information graph $\mathcal{G}$. $\hat{X}_e$ is said to be a unicast message if $\hat{X}_e$ is wanted by one receiver. Suppose that the maximal unicast acyclic subgraph of $\mathcal{G}$ consists of unicast messages $\bold{\hat{X}}_{E}$ and the receivers who want one of $\bold{\hat{X}}_{E}$, that is, $\{R_{e} | e\in E\}$. In such a case, the validity of the corresponding network structure is not guaranteed but for the directed acyclic network structure, it is guaranteed when we convert $\mathcal{G}$ to the corresponding network structure as in the previous section. Next, let $\bold{\hat{X}}_S$ and $\{R_t | t\in T\}$ be the remaining messages and the remaining receivers, respectively. Subsequently, we modify $\mathcal{G}$ to $\mathcal{G}^{\prime}$ and determine $\bold{\hat{X}}_{E^\prime}$ in $\mathcal{G}^{\prime}$ in order to ensure the validity of the corresponding network structure. Specifically, we determine $\bold{\hat{X}}_{E^\prime}$ by adding several messages to $\bold{\hat{X}}_{E}$ and determine $\{R_{e^\prime} | e^\prime\in E^\prime\}$ by adding a number of receivers to $\{R_{e} | e\in E\}$. Accordingly, some of the corresponding links of these added receivers are referred to as duplicated links, as explained later. Before choosing $\bold{\hat{X}}_{E^\prime}$, we classify problematic cases based on the outgoing edges of the receivers in $\mathcal{G}$, which should be modified to obtain $\mathcal{G}^\prime$.

There are six problematic cases based on message nodes and four problematic cases based on receiver nodes for the modification of the side information graph as in the following claim.

\vspace{2mm}
\begin{claim}
The problematic cases based on the outgoing edges of receivers in a side information graph $\mathcal{G}$ can be classified into the following ten cases, which should be modified to make the modified side information graph $\mathcal{G}^\prime$.

\begin{enumerate}
\item $\hat{X}_s$ has one incoming edge from $\{R_t | t\in T\}$ for $s\in S$.
\item $\hat{X}_s$ has more than one incoming edge from $\{R_t | t\in T\}$ for $s\in S$.
\item $\hat{X}_s$ has one incoming edge from $\{R_{e^\prime} | e^\prime\in E\}$ and one incoming edge from $\{R_t | t\in T\}$ for $s\in S$. 
\item $\hat{X}_e$ has one incoming edge from $\{R_{e^\prime} | e^\prime\in E\}$ and one incoming edge from $\{R_t | t\in T\}$ for $e\in E$. 
\item $\hat{X}_e$ has more than one incoming edge from $\{R_{e^\prime} | e^\prime\in E\}$ for $e\in E$. 
\item $\hat{X}_e$ has more than one incoming edge from $\{R_t | t\in T\}$ for $e\in E$.
\item $R_e$ has one outgoing edge to $\{\hat{X}_s | s\in S\}$ and one outgoing edge to $\{\hat{X}_{e^{\prime}} | e^\prime\in E\}$ for $e\in E$.
\item $R_t$ has one outgoing edge to $\{\hat{X}_s | s\in S\}$ for $t\in T$.
\item $R_t$ has more than one outgoing edge to $\{\hat{X}_s | s\in S\}$ for $t\in T$.
\item $R_t$ has one outgoing edge to $\{\hat{X}_s | s\in S\}$ and one outgoing edge to $\{\hat{X}_{e^{\prime}} | e^\prime\in E\}$ for $t\in T$.
\end{enumerate}
\label{claim:cases}
\end{claim}

\begin{figure}[t]
\centering
\subfigure[]{\includegraphics[scale=0.3]{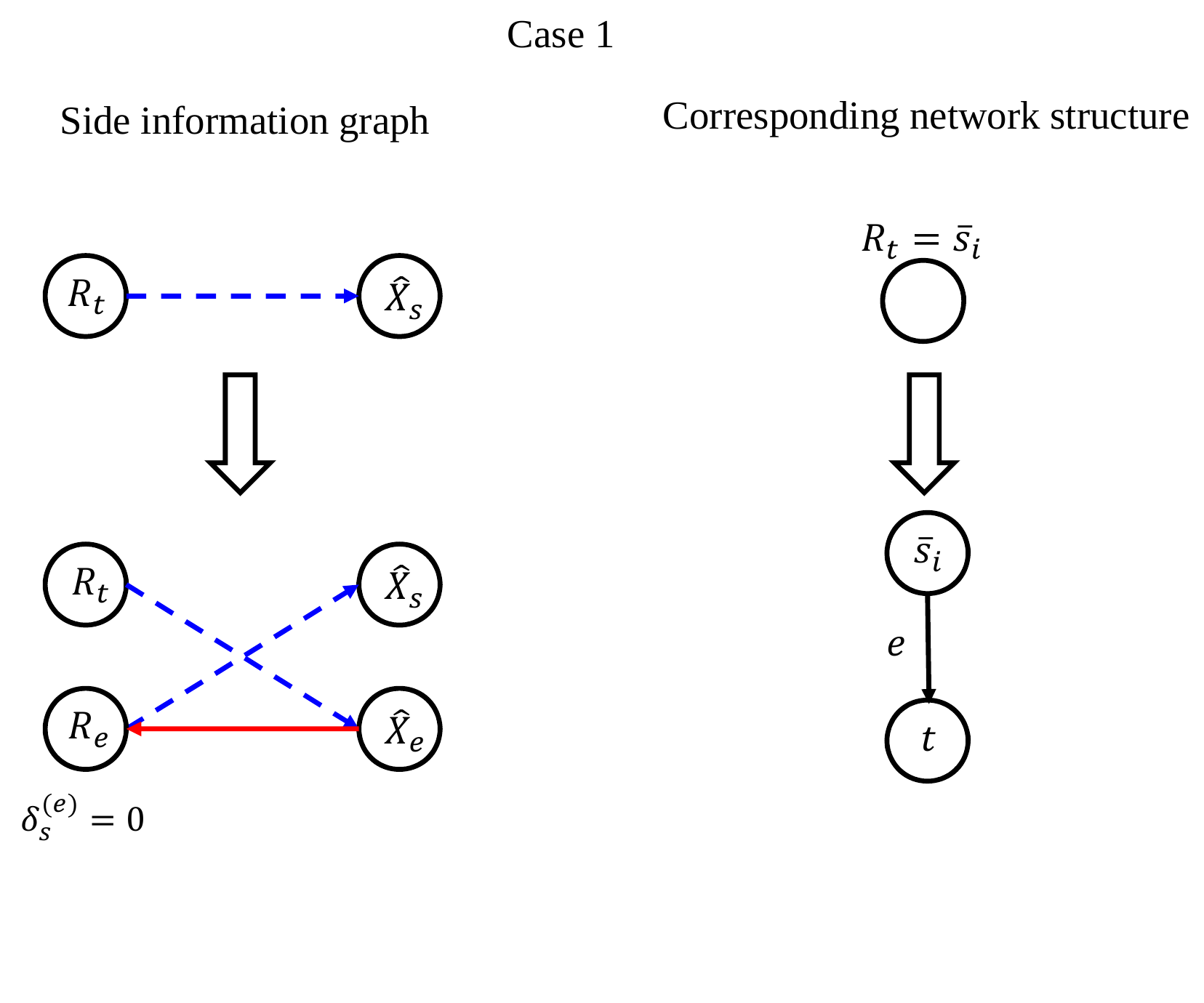}
\label{fig:fig2}}
\subfigure[]{\includegraphics[scale=0.3]{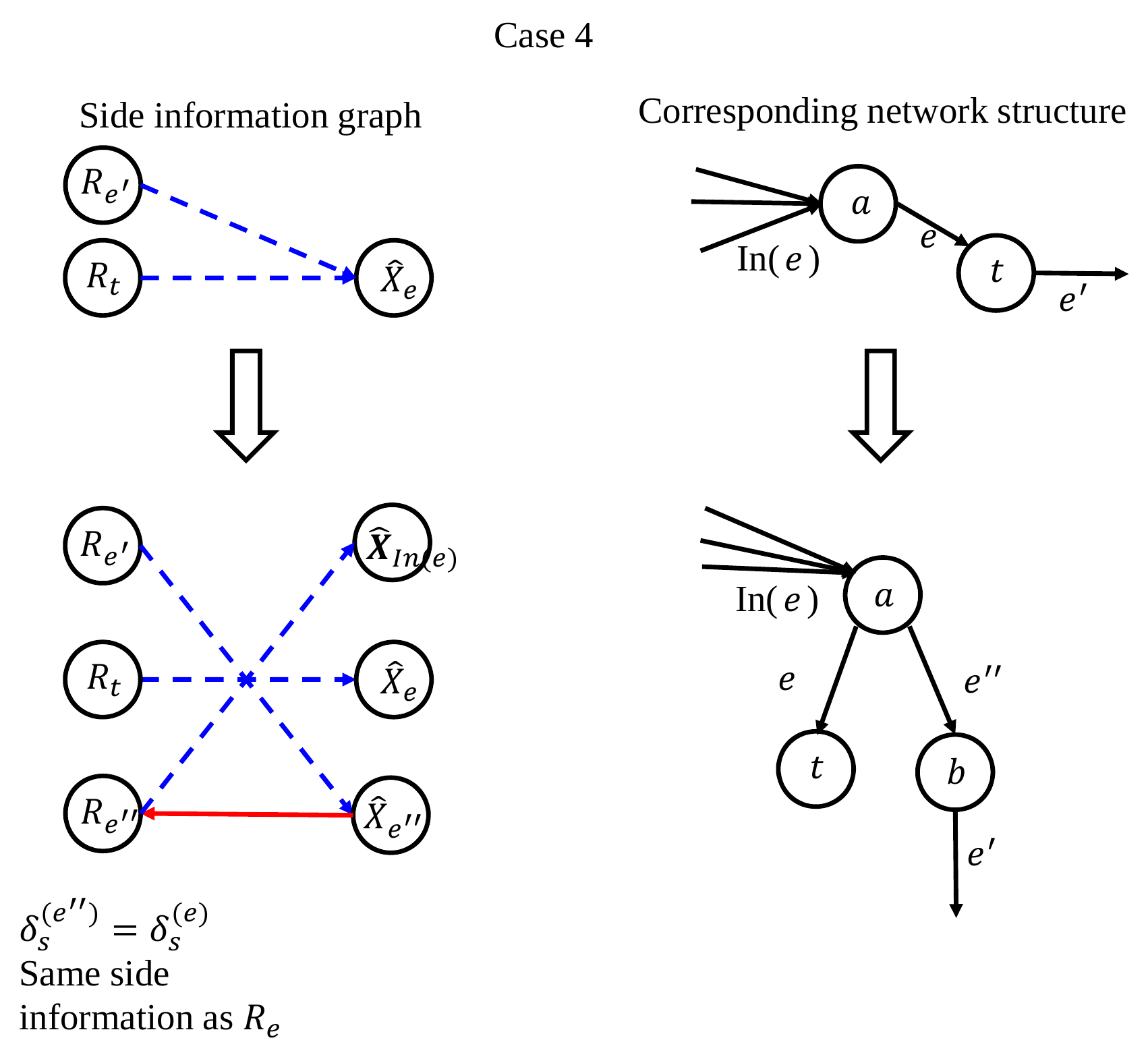}
\label{fig:fig3}}
\subfigure[]{\includegraphics[scale=0.3]{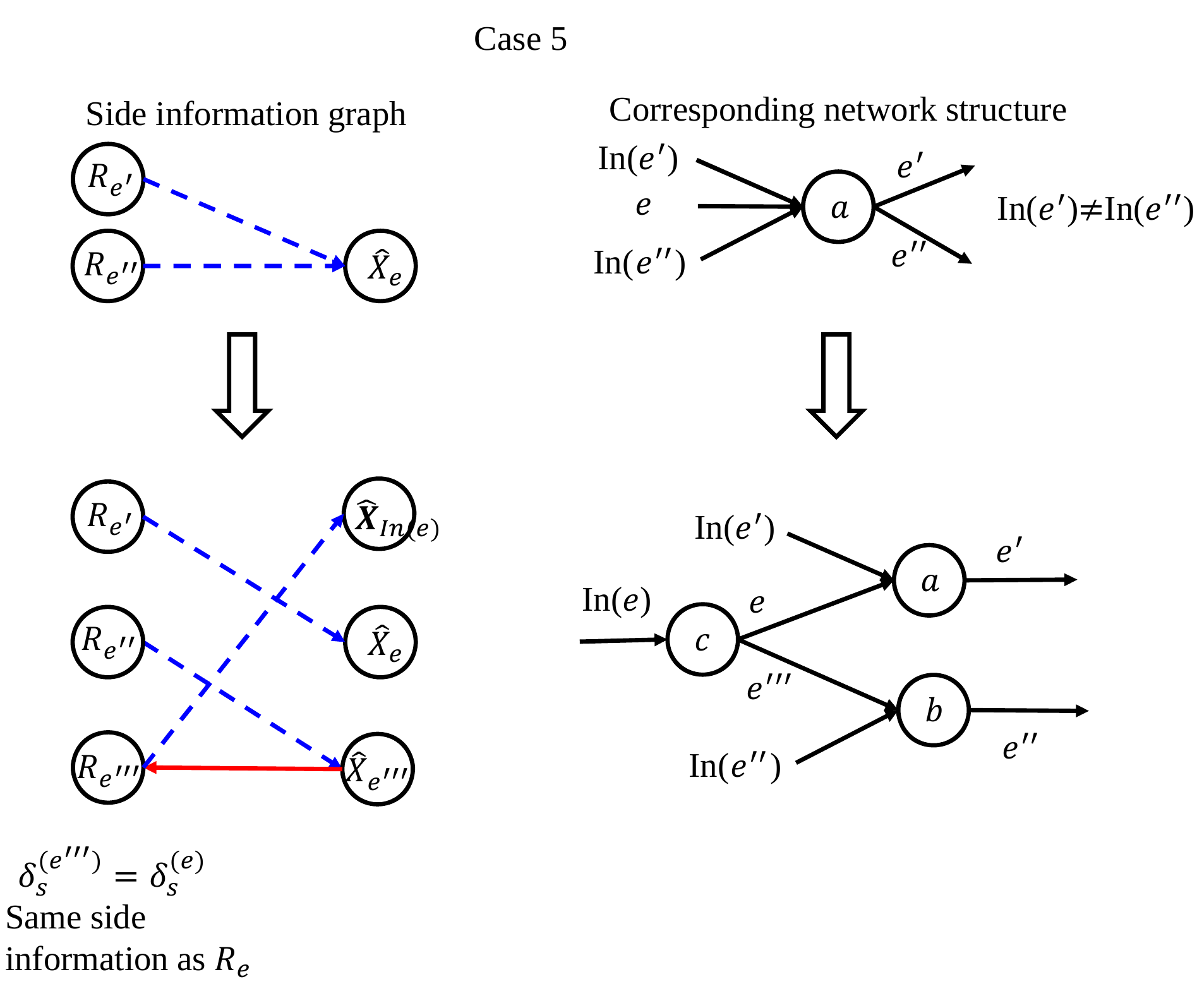}
\label{fig:fig4}}
\subfigure[]{\includegraphics[scale=0.3]{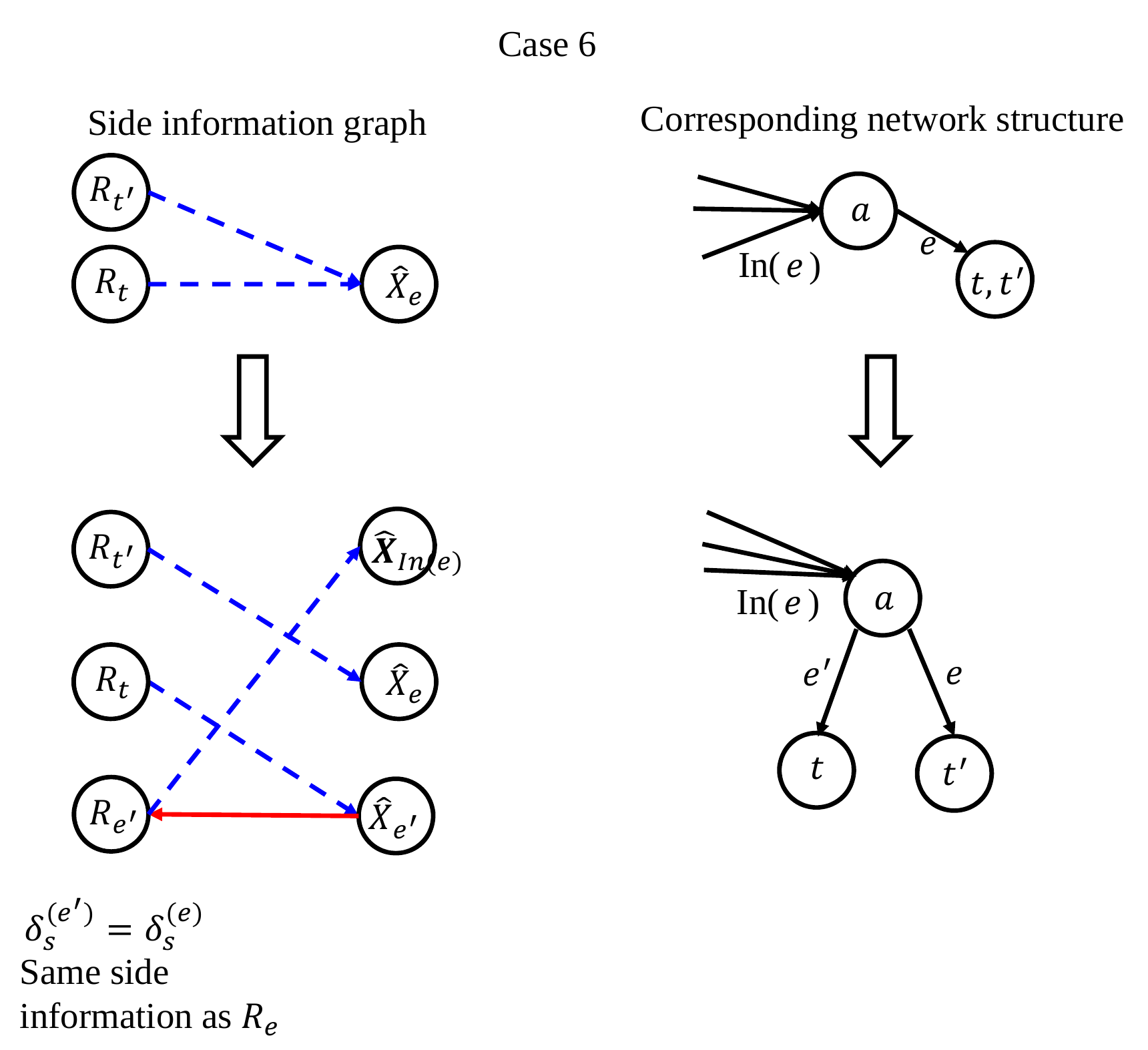}
\label{fig:fig5}}
\subfigure[]{\includegraphics[scale=0.3]{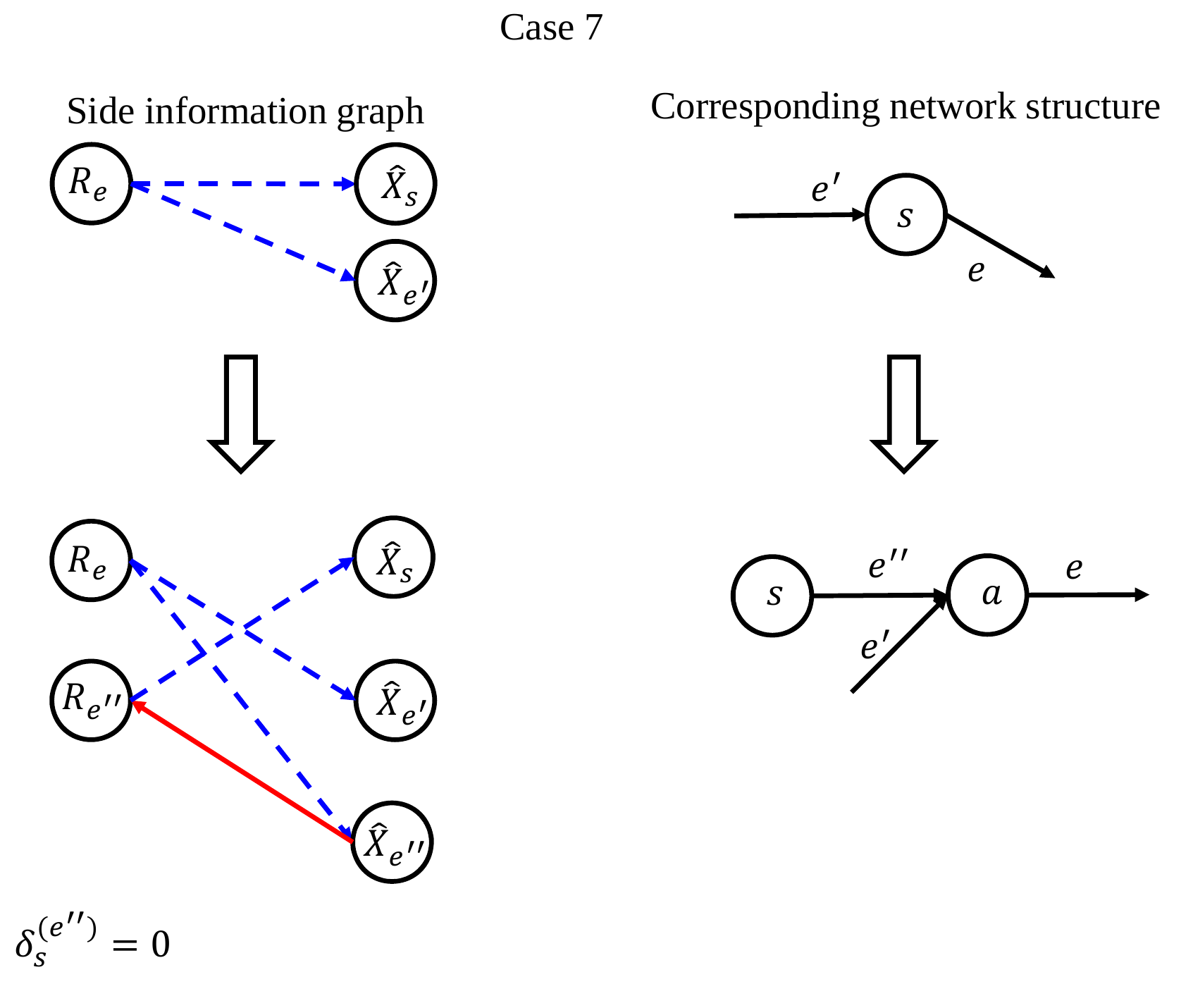}
\label{fig:fig6}}

\caption{Modifications of the problematic cases in Claim \ref{claim:cases}: (a) Case 1). (b) Case 4). (c) Case 5). (d) Case 6). (e) Case 7).} 
\label{fig:modification}
\end{figure}

In contrast to the ten problematic cases in Claim \ref{claim:cases}, there are ten cases which do not need to be modified for $\mathcal{G}^\prime$ as:
\begin{enumerate}
\item $\hat{X}_e$ has one incoming edge from $\{R_{e^\prime} | e^\prime\in E\}$ for $e\in E$. 
\item $\hat{X}_e$ has one incoming edge from $\{R_t | t\in T\}$ for $e\in E$.
\item $\hat{X}_s$ has one incoming edge from $\{R_{e^\prime} | e^\prime\in E\}$ for $s\in S$.
\item $\hat{X}_s$ has more than one incoming edge from $\{R_{e^\prime} | e^\prime\in E\}$ for $s\in S$.
\item $R_e$ has one outgoing edge to $\{\hat{X}_s | s\in S\}$ for $e\in E$.
\item $R_e$ has more than one outgoing edge to $\{\hat{X}_s | s\in S\}$ for $e\in E$.
\item $R_e$ has one outgoing edge to $\{\hat{X}_{e^{\prime}} | e^\prime\in E\}$ for $e\in E$.
\item $R_e$ has more than one outgoing edge to $\{\hat{X}_{e^{\prime}} | e^\prime\in E\}$ for $e\in E$.
\item $R_t$ has one outgoing edge to $\{\hat{X}_{e^{\prime}} | e^\prime\in E\}$ for $t\in T$.
\item $R_t$ has more than one outgoing edge to $\{\hat{X}_{e^{\prime}} | e^\prime\in E\}$ for $t\in T$.
\end{enumerate}

For example, the above case 1) can be described as $e^\prime$ being an incoming edge of $e$, which does not violate the network structure. Thus, the case 1) does not need to be modified.

At this point, we suggest how to modify the ten problematic cases in Claim \ref{claim:cases} so that the corresponding network coding instance is valid as in Fig. \ref{fig:modification}.

The case 1) in Claim \ref{claim:cases} is described as $R_t$ having $\hat{X}_s$ as side information for $t\in T$ and $s\in S$, implying that the terminal and the source are identical. To address this, we add a new link-related receiver ${R}_e$ having $\hat{X}_s$ as side information with $\delta_s^{(e)}=0$ and wanting the corresponding message $\hat{X}_e$. In addition, we delete the incoming edge of $\hat{X}_s$ from $R_t$ and add a new edge from $R_t$ to $\hat{X}_e$, after which we have the corresponding network structure as shown in Fig. \ref{fig:fig2}. The cases 2), 3), 8), 9), and 10) can be solved similarly to the case 1).

The case 4) indicates that the terminal node is the intermediate node, that is, $R_t$ and $R_{e^\prime}$ have $\hat{X}_e$ as side information. This can be modified by adding a new link-related receiver $R_{e^{\prime\prime}}$ having side information identical to that of $R_{e}$ with the identical $\delta_s^{(e^{\prime\prime})}=\delta_s^{(e)}$ and the corresponding message $\hat{X}_{e^{\prime\prime}}$, where $e^{\prime\prime}$ is a duplicated link of $e$. We delete the edge from $R_{e^\prime}$ to $\hat{X}_e$ and add a new edge from $R_{e^\prime}$ to $\hat{X}_{e^{\prime\prime}}$ and thus we have the corresponding network structure as shown in Fig. \ref{fig:fig3}. The case 5) can be a problem when two receivers with $\hat{X}_e$ as the side information have different side information as in Fig. \ref{fig:fig4}. This situation can be modified by a method similar to that in the case 4).

The case 6) is described as one in which $R_t$ and $R_{t^\prime}$ have $\hat{X}_e$ as side information, which means that the terminals $t$ and $t^\prime$ are identical. This situation can be modified by adding a new link-related receiver ${R}_{e^{\prime}}$ having the side information identical to that of $R_e$ with the same $\delta_s^{(e^{\prime})}=\delta_s^{(e)}$ and the corresponding message $\hat{X}_{e^\prime}$, where $e^{\prime}$ is a duplicated link of $e$. We delete the edge from $R_t$ to $\hat{X}_e$ and add a new edge from $R_t$ to $\hat{X}_{e^\prime}$, after which we have the corresponding network structure as shown in Fig. \ref{fig:fig5}. 

The case 7) indicates that the source node is the intermediate node, that is, $R_e$ has $\hat{X}_s$ and $\hat{X}_{e^{\prime}}$ as side information. This can be modified by adding a new link-related receiver $R_{e^{\prime\prime}}$ having $\hat{X}_s$ as side information with $\delta_s^{(e^{\prime\prime})}=0$ and the corresponding message $\hat{X}_{e^{\prime\prime}}$. We delete the edge from $R_e$ to $\hat{X}_s$ and add a new edge from $R_e$ to $\hat{X}_{e^{\prime\prime}}$. Accordingly, we have the corresponding network structure as shown in Fig. \ref{fig:fig6}.

By solving the above problematic cases and modifying $\mathcal{G}$ with $\bm{\delta}_s$ to $\mathcal{G}^{\prime}$ with $\bm{\delta}_s^\prime$, the valid corresponding network coding instance can be derived from any index coding instance.
Once the corresponding network coding instance is derived, we show the equivalence between an NCLE and an ICSIE for a given index coding instance as in the following theorem.

\vspace{2mm}
\begin{theorem}
For a given side information graph $\mathcal{G}$ with $\bm{\delta}_s$, a $(\bm{\delta}_s^{\prime},\mathcal{G}^{\prime})$-ICSIE with codelength $|E^\prime|$ exists if and only if the corresponding $(\bm{\delta},\mathbb{G})$-network code with link errors exists. 
\begin{proof}
By solving the problematic cases in Claim \ref{claim:cases}, we can determine $\{R_{e^\prime} | e\in E^{\prime}\}$ and $\bold{\hat{X}}_{E^\prime}$. Since we can have the valid network coding instance, a $(\bm{\delta}_s^{\prime},\mathcal{G}^{\prime})$-ICSIE with the codelength $|E^\prime|$ exists if and only if the corresponding $(\bm{\delta},\mathbb{G})$-network code with link errors exists by Theorem \ref{theorem:equ}.
\end{proof}
\label{theorem:indexequ1}
\end{theorem}
\vspace{2mm}

Our main concern is the equivalence between an index code for $\mathcal{G}$ and a network code for $\mathbb{G}$. However, Theorem \ref{theorem:indexequ1} shows the equivalence between an index code for $\mathcal{G}^{\prime}$ and a network code for $\mathbb{G}$. Thus, we have the following corollary.

\vspace{2mm}
\begin{corollary}
A $(\bm{\delta}_s,\mathcal{G})$-ICSIE with the codelength $|E|$ exists if and only if the corresponding $(\bm{\delta},\mathbb{G})$-network code with link errors exists, where each encoding function of the duplicated links is a function of each encoding function of the original links in the network code. 
\begin{proof}
Using the problematic cases in Claim \ref{claim:cases}, we prove the corollary, where the same notations for the above cases are used.

Necessity: Assume that a $(\bm{\delta}_s,\mathcal{G})$-ICSIE with codelength $|E|$ exists. For the case 1) in Claim \ref{claim:cases}, we need one more transmission $\hat{X}_e+\hat{X}_s$. Then, $R_t$ can still obtain what $R_t$ wants because $R_t$ can obtain $\hat{X}_{\tilde{s}}$ from $\hat{X}_e+\hat{X}_s-\hat{X}_{\tilde{e}}$, which is the same situation as before. Trivially, ${R}_e$ can obtain $\hat{X}_e$ from $\hat{X}_e+\hat{X}_s-\hat{X}_s$. For the case 4), we also need one more transmission $\hat{X}_e+\hat{X}_{e^{\prime\prime}}$. Then, $R_{e^{\prime}}$ can still obtain what $R_{e^{\prime}}$ wants because $R_{e^{\prime}}$ can obtain $\hat{X}_{\tilde{e}}$ from $\hat{X}_e+\hat{X}_{e^{\prime\prime}}-\hat{X}_{\tilde{e^{\prime\prime}}}$. $R_{e^{\prime\prime}}$ can easily obtain $\hat{X}_{e^{\prime\prime}}$ from $\hat{X}_e+\hat{X}_{e^{\prime\prime}}-\hat{X}_e$ because $R_{e^{\prime\prime}}$ can recover $\hat{X}_e$. For the case 5), we need one more transmission $\hat{X}_{e}+\hat{X}_{e^{\prime\prime\prime}}$. Then, $R_{e^{\prime\prime}}$ can still obtain what $R_{e^{\prime\prime}}$ wants because $R_{e^{\prime\prime}}$ can obtain $\hat{X}_{\tilde{e}}$ from $\hat{X}_{e}+\hat{X}_{e^{\prime\prime\prime}}-\hat{X}_{\tilde{e^{\prime\prime\prime}}}$. Clearly, $R_{e^{\prime\prime\prime}}$ can obtain $\hat{X}_{e^{\prime\prime\prime}}$ from $\hat{X}_e+\hat{X}_{e^{\prime\prime\prime}}-\hat{X}_e$ because $R_{e^{\prime\prime\prime}}$ can recover $\hat{X}_e$. For the case 6), we need one more transmission $\hat{X}_{e^\prime}+\hat{X}_e$. Then, $R_t$ can still obtain what $R_t$ wants because $R_t$ can obtain $\hat{X}_{\tilde{e}}$ from $\hat{X}_{e^\prime}+\hat{X}_e-\hat{X}_{\tilde{e^\prime}}$. $R_{e^\prime}$ can easily recover $\hat{X}_{e^\prime}$ from $\hat{X}_{e^\prime}+\hat{X}_e-\hat{X}_e$ because $R_{e^\prime}$ can recover $\hat{X}_e$. For the case 7), we also need one more transmission $\hat{X}_s+\hat{X}_{e^{\prime\prime}}$. Then, $R_e$ can obtain $\hat{X}_e$ because $R_e$ can obtain $\hat{X}_{\tilde{s}}$ from $\hat{X}_s+\hat{X}_{e^{\prime\prime}}-\hat{X}_{\tilde{e^{\prime\prime}}}$, which is identical to the earlier situation. ${R}_{e^{\prime\prime}}$ can easily obtain $\hat{X}_{e^{\prime\prime}}$ from $\hat{X}_s+\hat{X}_{e^{\prime\prime}}-\hat{X}_s$. Thus, a $(\bm{\delta}_s^{\prime},\mathcal{G}^{\prime})$-ICSIE with codelength $|E^\prime|$ exists if a $(\bm{\delta}_s,\mathcal{G})$-ICSIE with codelength $|E|$ exists using additional transmissions as described above because the number of additional transmissions is $|E^\prime|-|E|$ and thus the corresponding $(\bm{\delta},\mathbb{G})$-NCLE exists by Theorem \ref{theorem:indexequ1}. Since the encoding functions of the corresponding network code are defined by the decoding functions of a given index code and each decoding function of the added receivers is a function of each decoding function of the original receivers, each encoding function of duplicated links is a function of each encoding function of the original links in the network code. 

Sufficiency: Suppose that a $(\bm{\delta},\mathbb{G})$-NCLE exists. Then, we have a $(\bm{\delta}_s^{\prime},\mathcal{G}^{\prime})$-ICSIE with codelength $|E^\prime|$, that is, $\bold{\hat{X}}_B=(\hat{X}_B(e): e\in E^\prime)$, where $\hat{X}_B(e)=\hat{X}_e+\bar{F}_e(\hat{X}_1,...,\hat{X}_{|S|})$. In fact, selecting $|E|$ components of the given index code is sufficient for making a $(\bm{\delta}_s,\mathcal{G})$-ICSIE with codelength $|E|$. For the case 1), $R_t$ can obtain what $R_t$ wants even if we do not transmit $\hat{X}_e+\bar{F}_e(\hat{X}_1,...,\hat{X}_{|S|})$. $R_t$ simply needs $\bar{F}_e(\hat{X}_s)$ related to $e$. Since $R_t$ has $\hat{X}_{\tilde{s}}$ as side information in $\mathcal{G}$, $R_t$ can calculate $\bar{F}_e(\hat{X}_s)$, which may be erroneous. The case 7) is derived similarly to how the case 1) was derived. For the case 4), showing that $R_{e^\prime}$ can obtain $\hat{X}_{e^\prime}$ even though we do not transmit $\hat{X}_{e^{\prime\prime}}+\bar{F}_{e^{\prime\prime}}(\hat{X}_1,...,\hat{X}_{|S|})$ is sufficient.
Since $R_{e^\prime}$ has $\hat{X}_{\tilde{e}}$ as side information in $\mathcal{G}$, $R_{e^\prime}$ can have $\bar{F}_e(\hat{X}_1,...,\hat{X}_{|S|})$, which may be erroneous. Thus, $R_{e^\prime}$ can obtain $\hat{X}_{e^\prime}$ if $\bar{F}_{e^{\prime\prime}}$ is a function of $\bar{F}_e$. The cases 5) and 6) are derived similarly to the case 4).
\end{proof}
\label{cor:indexequ2}
\end{corollary}
\vspace{2mm}

\begin{remark}
For a given index coding instance, we can make the corresponding network coding instance by introducing some links, referred to as duplicated links, which are in fact duplicated encoding functions. The duplicated links in the corresponding network coding instance can always be made depending on the original links but some of them can be made independent in the perspective of network coding. Thus, the sufficiency in Corollary \ref{cor:indexequ2} holds for the intended dependent situations, where each encoding function of the duplicated links is a function of each encoding function of the original links.  
\end{remark}
\vspace{2mm}

In \cite{FEQU}, the corresponding network coding instance can easily be converted from the original index coding instance. However, the given original network coding instance is different from the network coding instance re-converted from the corresponding index coding instance. That is, assume that for a given network coding structure $\mathbb{G}$ with $\bm{\delta}$, we have the corresponding side information graph $\mathcal{G}$ with $\bm{\delta}_s$. If we derive the corresponding network coding structure $\mathbb{G}^{\prime}$ from $\mathcal{G}$ using the method in \cite{FEQU}, $\mathbb{G}^{\prime}$ is always different from $\mathbb{G}$. Similarly, the same problem occurs for a given index coding instance. However, for a given network coding instance, we can make $\mathbb{G}=\mathbb{G}^{\prime}$ with $\bm{\delta}=\bm{\delta^\prime}$ using the proposed modification of the side information graph and ensure convertibility between the two codes when the encoding functions of the duplicated links are functions of the encoding functions of the original links in the corresponding network code. A similar approach can be applied to a problem for a given index coding instance.   

\vspace{2mm}
\begin{example}

\begin{figure}[t]
\centering
\subfigure[]{\includegraphics[scale=0.55]{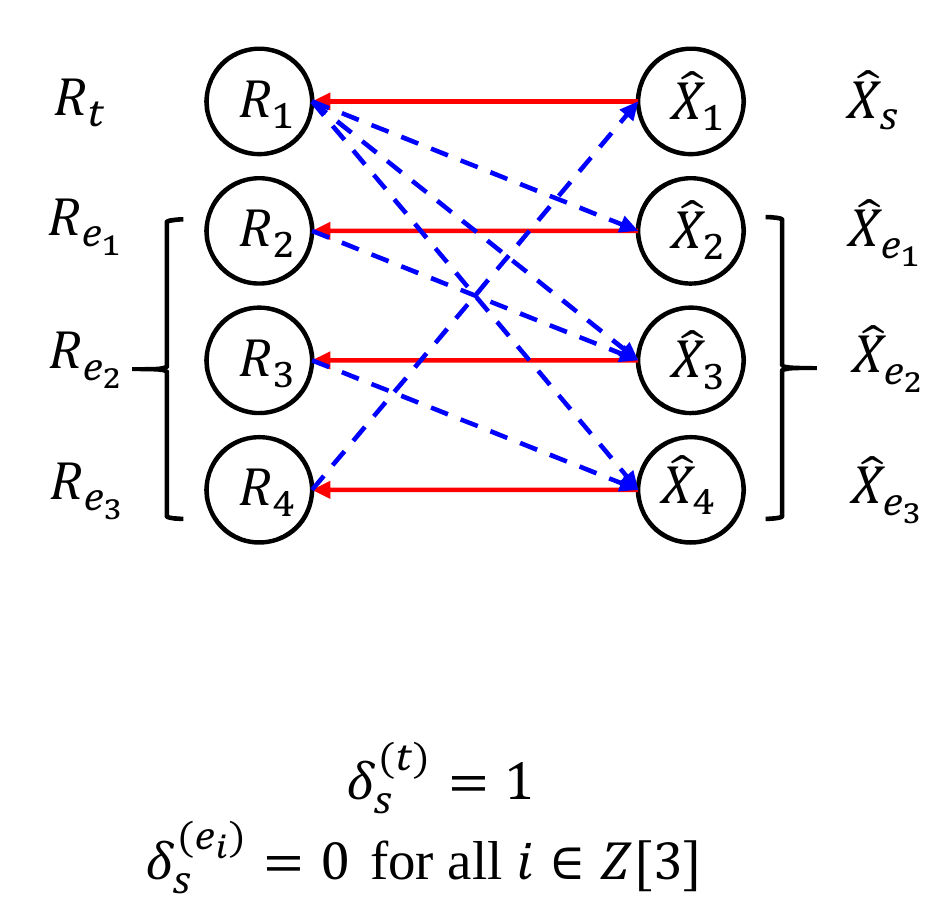}
\label{fig:fig8}}
\subfigure[]{\includegraphics[scale=0.55]{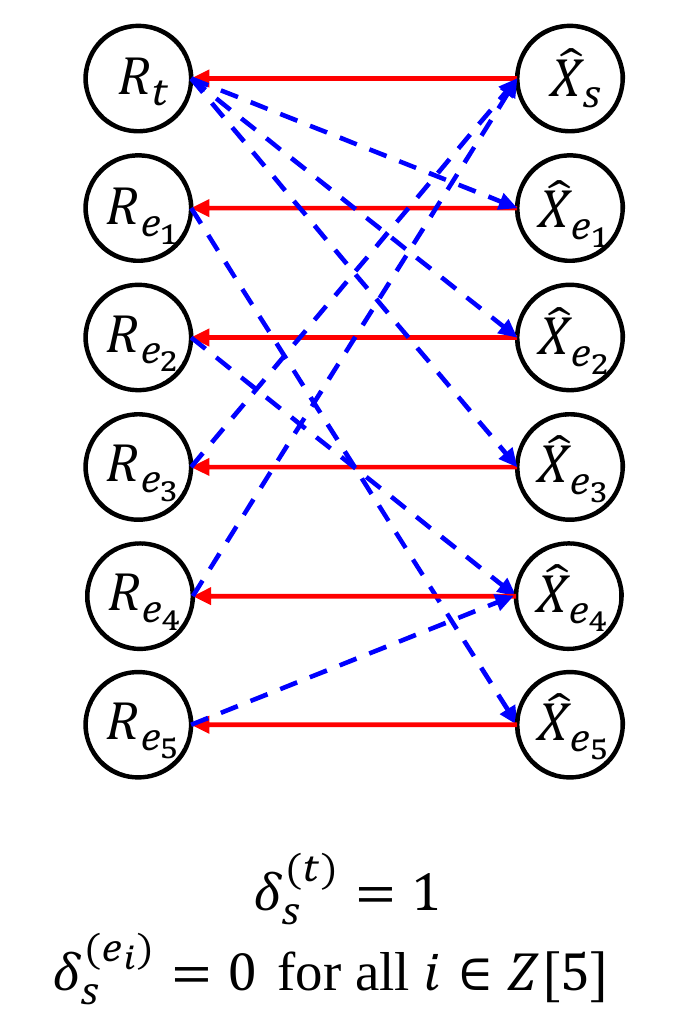}
\label{fig:fig9}}
\subfigure[]{\includegraphics[scale=0.55]{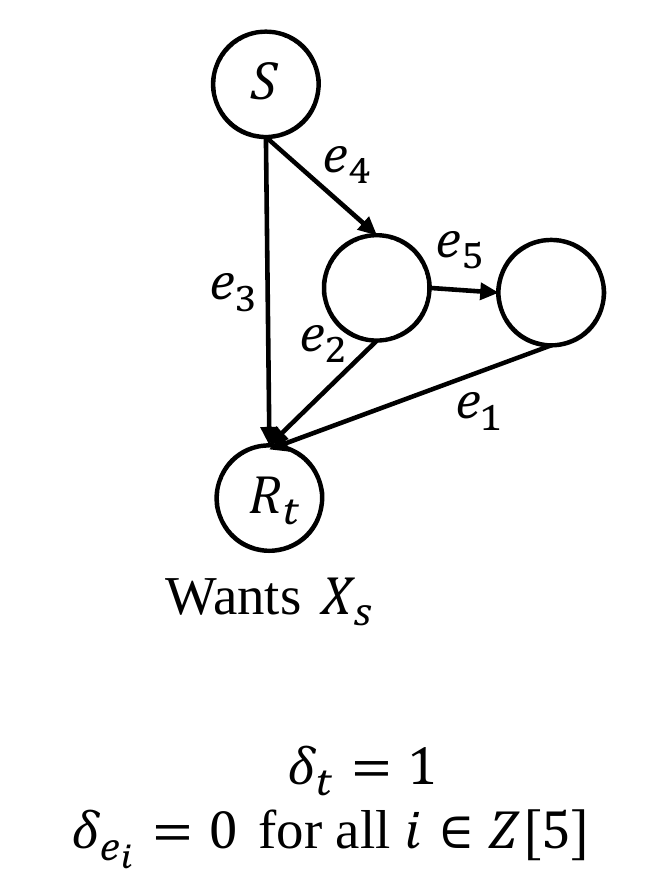}
\label{fig:fig10}}

\caption{The description of Example \ref{ex:given index}: (a) A given side information graph $\mathcal{G}$ with $\bm{\delta}_s$. (b) A modified side information graph $\mathcal{G}^{\prime}$ with $\bm{\delta}^{\prime}_s$. (c) The corresponding network coding instance $\mathbb{G}$.} 
\end{figure}

Suppose that a given side information graph $\mathcal{G}$ with $\bm{\delta}_s$ is given in Fig. \ref{fig:fig8}. Then, a modified side information graph $\mathcal{G}^\prime$ with $\bm{\delta}^{\prime}_s$ and the corresponding network coding instance are shown in Fig. \ref{fig:fig9} and Fig. \ref{fig:fig10}, respectively. We assume the field size $q=2$.

We first find the maximal unicast acyclic subgraph of $\mathcal{G}$ and determine $\bold{\hat{X}}_{E}$, $\hat{X}_s$, and the corresponding receivers as in Fig. \ref{fig:fig8}. Subsequently, we can make a modified side information graph by solving the case 4) in Claim \ref{claim:cases} twice. An index code for $\mathcal{G}$ with codelength $|E|=3$ is $(\hat{X}_s+\hat{X}_{e_1},\hat{X}_{e_1}+\hat{X}_{e_2},\hat{X}_{e_2}+\hat{X}_{e_3})$. Then, every receiver can recover what it wants. For example, $R_t$ can calculate $\hat{X}_s+\hat{X}_{e_1}$, $\hat{X}_s+\hat{X}_{e_2}$, and $\hat{X}_s+\hat{X}_{e_3}$ from the received codeword. Since $\delta_s^{(t)}=1$ and $R_t$ has $\hat{X}_{\tilde{e_1}}$, $\hat{X}_{\tilde{e_2}}$, and $\hat{X}_{\tilde{e_3}}$ as side information, subtracting them from $\hat{X}_s+\hat{X}_{e_1}$, $\hat{X}_s+\hat{X}_{e_2}$, and $\hat{X}_s+\hat{X}_{e_3}$, respectively results in the true symbol by majority decoding. With Theorem \ref{theorem:indexequ1} and Corollary \ref{cor:indexequ2}, we can find an index code for $\mathcal{G}^{\prime}$ as $(\hat{X}_s+\hat{X}_{e_1},\hat{X}_{e_1}+\hat{X}_{e_2},\hat{X}_{e_2}+\hat{X}_{e_3},\hat{X}_{e_3}+\hat{X}_{e_4},\hat{X}_{e_2}+\hat{X}_{e_5})$ and a network code for $\mathbb{G}$ as follows:

\begin{enumerate}
\item $X_{e_1}=\hat{D}_{e_1}(\bold{0},(X_{e^\prime} : e^\prime\in {\rm In}(e_1)))=0+0+X_{e_5}=X_{e_5}$
\item $X_{e_2}=0+0+X_{e_4}=X_{e_4}$
\item $X_{e_3}=0+0+0+X_{s}=X_s$
\item $X_{e_4}=0+0+0+0+X_s=X_s$
\item $X_{e_5}=0+0+0+X_{e_4}=X_{e_4}$
\item $D_t=\hat{D}_t(\bold{0},(\tilde{X}_{e^\prime} : e^\prime\in {\rm In}(t)))$
\end{enumerate}

By Claim \ref{claim:cases}, $e_4$ is a duplicated link of $e_3$ and $e_5$ is a duplicated link of $e_2$. It is clear that $X_{e_4}$ is a function of $X_{e_3}$ and $X_{e_5}$ is a function of $X_{e_2}$. Similarly, we can find an index code for $\mathcal{G}$ from a network code for $\mathbb{G}$ if each encoding function of the duplicated links is a function of each encoding function of the original links in the network code.

\label{ex:given index}
\end{example}

\vspace{2mm}
\begin{remark}

In general, a given side information graph can be converted to several distinct modified side information graphs but any modified side information graph can be re-converted to the original side information graph. Furthermore, there is a one-to-one correspondence between a modified index coding instance and the corresponding network coding instance. Thus, for a given index coding instance, there are several distinct corresponding network coding instances but each corresponding network coding instance can be re-converted to the original index coding instance.
\end{remark}
\vspace{2mm}

\section{Relationship of Some Properties}\label{Properties}

In this section, several properties of an NCLE are introduced using the properties of an ICSIE. Since the equivalence between an NCLE and an ICSIE is shown when either a network coding instance or an index coding instance is given, we can utilize the properties of an ICSIE to derive those of an NCLE. First, we introduce a property of an ICSIE in the following lemma, which is similar to that in \cite{ICSIE}.

\vspace{2mm}
\begin{lemma}
Suppose that a $(\bold{0},\bar{\mathcal{G}})$-IC problem is constructed by deleting any less than or equal to min$(2\delta_s^{(i)}, |\mathcal{X}_i|)$ outgoing edges from each receiver $R_i$ in a $(\bm{\delta}_s,\mathcal{G})$-ICSIE problem. That is, each receiver of $\bar{\mathcal{G}}$ has larger than or equal to max$(0, |\mathcal{X}_i|-2\delta_s^{(i)})$ side information symbols and then it becomes the conventional index coding problem. Then, $N_{\rm opt}^q(\bold{0},\bar{\mathcal{G}})\leq N_{\rm opt}^q(\bm{\delta}_s,\mathcal{G})$.

\begin{proof}
This is proved similarly to the method in \cite{ICSIE} and thus we omit it here.
\end{proof}
\label{lowerbound}
\end{lemma}
\vspace{2mm}

Lemma \ref{lowerbound} shows the relationship between the conventional index code and an ICSIE. Thus, we can infer that a property between the conventional network code with $\bm{\delta}=\bold{0}$ and an NCLE is derived by Lemma \ref{lowerbound} as in the following theorem.

\vspace{2mm}
\begin{theorem}
Let $E_v$ be a set of outgoing links of $v\in V$ and $\delta_v$ be ${\rm min}\{\delta_e | e\in E_v\}$. If a $(\bm{\delta},\mathbb{G})$-NCLE exists for a given network structure $\mathbb{G}$, there exists the conventional network code with $\bm{\delta}=\bold{0}$ after deleting arbitrary $2\delta_v$ links from ${\rm In}(v)$ for all $v\in V$ in $\mathbb{G}$.
\begin{proof}
Let $\bar{\mathcal{G}}$ be the side information graph of the corresponding index code of the conventional network code and $\mathcal{G}$ be the side information graph of the corresponding index code of a $(\bm{\delta},\mathbb{G})$-NCLE. Instead of deleting arbitrary $2\delta_v$ incoming links for all $v\in V\setminus\bar{S}$ in $\mathbb{G}$, we can consider these links as incoming links of dummy nodes with no outgoing link. Then, from Theorem \ref{theorem:equ} and Observation \ref{ob:codelength}, $N_{\rm{opt}}^q(\bold{0},\bar{\mathcal{G}})=|E|$ if the conventional network code is valid. From Lemma \ref{lowerbound}, $N_{\rm opt}^q(\bold{0},\bar{\mathcal{G}})\leq N_{\rm opt}^q(\bm{\delta}_s,\mathcal{G})$ and $N_{\rm opt}^q(\bm{\delta}_s,\mathcal{G})=|E|$ if the $(\bm{\delta},\mathbb{G})$-NCLE is feasible. If the conventional network code is not valid, $N_{\rm opt}^q(\bold{0},\bar{\mathcal{G}})>|E|$, which results in the fact that the $(\bm{\delta},\mathbb{G})$-NCLE is not feasible.
\end{proof}
\end{theorem}
\vspace{2mm}

Thus, we can find the conventional network code whenever we have an NCLE. Next, another property of an NCLE is introduced. Before showing it, we define an independent component of an index code.
 
 \vspace{2mm}
\begin{definition}
A component $\hat{X}_e$ in $\bold{\hat{X}}_E$ is said to be independent if fixing the value of $\hat{X}_e$ results in reduction of the code dimension by one.
\end{definition}
\vspace{2mm}

\begin{remark}
Independent components in the corresponding index coding instance are always in $\bold{\hat{X}}_E$ because a network structure is directed acyclic and $N_{\rm opt}^q(\bm{\delta}_s,\mathcal{G})=|E|$.
\end{remark}
\vspace{2mm}

Now, we introduce a property related to the redundant links of a network code. From the equivalence between a network code and an index code, we can infer that redundant links may be related to some properties of an index code as shown in the following theorem.

\vspace{2mm}
\begin{theorem}
Redundant links in a network code are equivalent to independent components in $\bold{\hat{X}}_E$ of the corresponding index code.
\begin{proof}
Necessity: If $e$ is a redundant link in the given network code, removing $e$ does not affect the feasibility of the given network code. If we remove $e$, the corresponding index code should have codelength $|E|-1$ by Theorem \ref{theorem:equ}. It means that removing $R_e$ and $\hat{X}_e$ from the corresponding index coding problem results in reduction of the code dimension by one.

Sufficiency: If $\hat{X}_e$ is an independent component, fixing its value causes reduction of the code dimension by one and this index code is feasible by Theorem \ref{encoding}. Thus, we can say that $e$ is a redundant link.
\end{proof}
\end{theorem}
\vspace{2mm}

\section{Conclusions}\label{Sec:Conclusion}

In this paper, a new equivalence between an NCLE and an ICSIE was proposed. In order to provide the equivalence between an NCLE and an ICSIE, we considered a new type of a network code, referred to as a $(\bm{\delta},\mathbb{G})$-NCLE, where the intermediate nodes can resolve incoming errors.

First, we showed the equivalence between a $(\bm{\delta},\mathbb{G})$-NCLE and a $(\bm{\delta}_s,\mathcal{G})$-ICSIE for a given network coding instance with $\mathbb{G}$ and $\bm{\delta}$. We also showed that the corresponding side information graph does not need the receiver $\hat{t}_{\rm all}$, which is contained in the previous models \cite{EQU}, \cite{FEQU}. 

In addition to the equivalence between an NCLE and an ICSIE for a given network coding instance, their equivalence was also derived for a given index coding instance. For a given side information graph $\mathcal{G}$ with $\bm{\delta}_s$, we derived the corresponding network coding instance with $\mathbb{G}$ and $\bm{\delta}$ by modifying $\mathcal{G}$ with $\bm{\delta}_s$ to $\mathcal{G}^{\prime}$ with $\bm{\delta^\prime}_s$. With the proposed method of modifying $\mathcal{G}$, we showed an equivalence between a $(\bm{\delta},\mathbb{G})$-NCLE and a $(\bm{\delta}_s,\mathcal{G})$-ICSIE for a given index coding instance if a pair of encoding functions of the original link and the duplicated link are functionally related.

Finally, several properties of a $(\bm{\delta},\mathbb{G})$-NCLE were derived from the properties of a $(\bm{\delta}_s,\mathcal{G})$-ICSIE using their equivalence relationship.

\end{document}